\documentclass{article}
\usepackage[T1]{fontenc}
\usepackage{iclr2027_conference,times}
\usepackage{amsmath,amssymb,graphicx,booktabs}
\usepackage{colortbl}
\usepackage{placeins}
\usepackage{flafter}
\usepackage{capt-of}
\usepackage{booktabs}
\usepackage{tabularx}
\usepackage{amsmath}
\usepackage{amssymb}
\usepackage{algorithm}
\usepackage{algpseudocode}

\definecolor{oursrow}{HTML}{E8F2FA}
\definecolor{rankfirst}{HTML}{FFD9D9}
\definecolor{ranksecond}{HTML}{DEFADE}
\definecolor{rankthird}{HTML}{DFDFFF}
\usepackage{hyperref}
\hypersetup{hidelinks}
\usepackage{url}
\title{UNBIND: UNlearning By INference-time \\Directional Steering for Code LLMs}
\author{
\textbf{Zhengyang Shan, Jiayun Xin, Yanjun Lin, Xu Qian, Zhiang Liu,} \\
\textbf{Minghui Xu, Yue Zhang, Qin Hu, Kun Li, Xiuzhen Cheng} \\
Department of Computer Science and Technology \\
Shandong University \\
Qingdao, China \\
\texttt{\{202520912,202535317,202635420,202515294,202635451\}@mail.sdu.edu.cn} \\
\texttt{mhxu@sdu.edu.cn, zyueinfosec@sdu.edu.cn} \\
\texttt{qinhu@sdu.edu.cn, kunli@sdu.edu.cn} \\
\texttt{xzcheng@sdu.edu.cn}
}
\iclrfinalcopy
\begin{document}
\maketitle
\lhead{}
\begin{center}
\begin{minipage}{\linewidth}
\centering
\includegraphics[width=\linewidth]{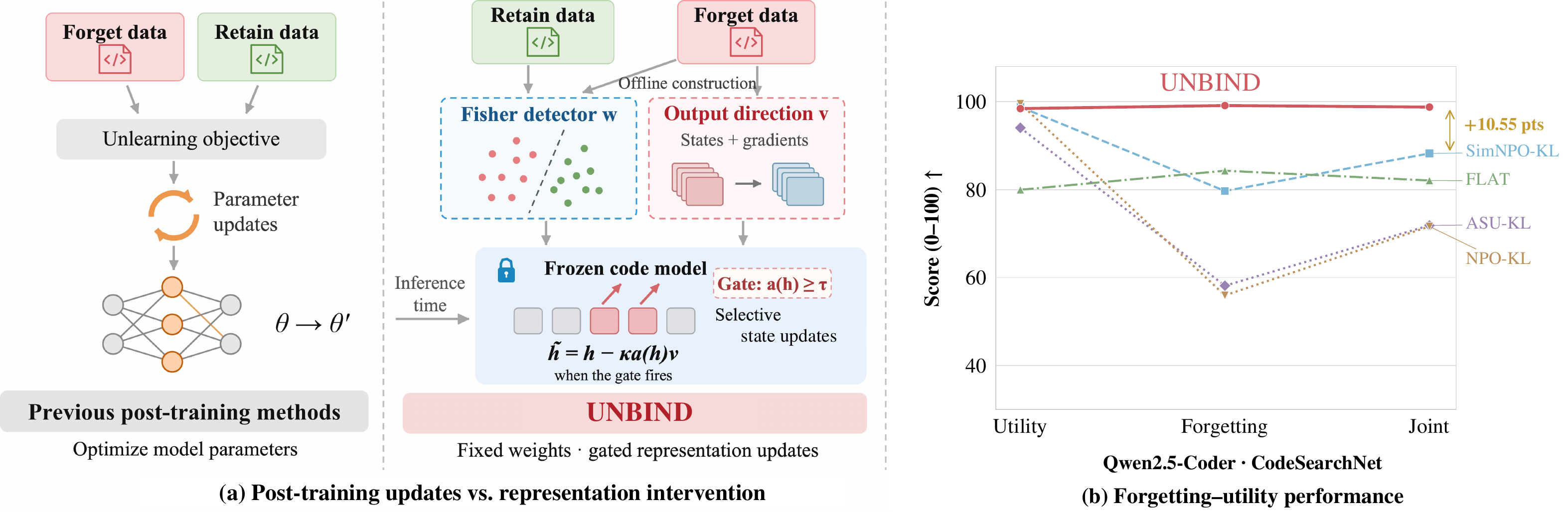}
\captionof{figure}{\textbf{Overview of UNBIND.}
(a) Unlike post-training unlearning methods that update model parameters, UNBIND keeps model weights fixed and intervenes on hidden states at inference time.
(b) UNBIND achieves strong forgetting while preserving model utility, outperforming representative post-training baselines.
}
\label{fig:overview}
\end{minipage}
\end{center}
\begin{abstract}
Code large language models acquire programming capabilities from large code corpora, but can also memorize implementations that later require removal. Code unlearning is needed to control their continued reproduction when copyright or security concerns arise. However, targeted and retained code share computational patterns, creating a tension between forgetting specific implementations and preserving general programming ability. We propose \textbf{UNBIND}, a code unlearning framework that separately considers which hidden states correspond to the target code and how to suppress its reproduction. By constructing separate directions for these objectives, UNBIND achieves selective unlearning at inference time while keeping model weights fixed. Our evaluation covers fourteen baselines across two code models and two corpora. UNBIND achieves the highest joint forgetting and utility score in every setting. It reduces target code reproduction by 97.3\% to 99.1\% as measured by F-BLEU, with at most two fewer HumanEval+ and six fewer MBPP+ problems solved than the original models. In repeated extraction tests under a fixed budget, the number of targets yielding exact spans of at least 50 tokens falls from 188--262 to 0--2 out of 300 per setting. No extracted span reaches 100 tokens, and the mean best recovery ratio ranges from 0.43\% to 6.45\%. Multilingual and related-code evaluations further show effective forgetting with limited impact on useful programming capabilities, supporting UNBIND as a practical approach to selective code unlearning.
\end{abstract}

\vspace{-2.0em}
\section{Introduction}
Code large language models support program completion and generation by learning from large code corpora~\citep{li2023starcoder,roziere2023codellama,hui2024qwen25coder}. They can also memorize and reproduce specific implementations~\citep{carlini2021extracting,carlini2023quantifying}. Some may later require removal because of copyright or security concerns~\citep{jiang2026prod}. Removing code from the corpus does not change a trained model. Code unlearning is therefore needed to reduce target reproduction while preserving general programming ability.

The key challenge is that low target reproduction can result from degraded programming ability rather than selective forgetting. Target and retained code share computational patterns, so representations that help reproduce a particular implementation may also support programs that should remain unaffected. For example, suppressing a control-flow pattern used by the target may interfere with other implementations that rely on the same pattern. A representation associated with the target is therefore not necessarily specific to it. Stronger suppression can reduce target reproduction while also weakening predictions needed to generate correct code elsewhere. Restricting the magnitude or scope of these changes can preserve utility, but may leave the predictions responsible for target reproduction insufficiently altered. This makes it difficult to improve forgetting through suppression strength alone. The changes must effectively weaken target predictions while limiting their impact on computations that support other programs. Several baselines in our evaluation illustrate this tradeoff, achieving low reproduction with substantial losses on independent programming tasks.

To address this tradeoff, we separate recognizing target-related representations from suppressing target predictions, since a direction that distinguishes forget from retain states need not suppress target predictions effectively. We introduce \textbf{UNBIND}: UNlearning By INference-time Directional Steering of Code Representations, as outlined in Figure~\ref{fig:overview}. At a selected Transformer layer, a Fisher detector identifies target-related representations using forget and retain hidden states. A separate gradient-guided output direction uses forget states and their target loss gradients to suppress target predictions. During inference, a calibrated gate selects positions to update along this direction, with magnitudes determined by detector activation. Model and adapter weights remain fixed.

Our evaluation covers fourteen baselines across two code models and two corpora. UNBIND achieves the highest joint forgetting and utility score in every setting. It reduces F-BLEU by 97.3\% to 99.1\%, with at most two fewer HumanEval+ and six fewer MBPP+ problems solved than the original models. In repeated extraction tests under a fixed budget, targets yielding contiguous exact spans of at least 50 tokens fall from 188--262 to 0--2 out of 300 per setting. No span reaches 100 tokens, and the mean best recovery ratio ranges from 0.43\% to 6.45\%.

Our contributions are summarized as follows:
\vspace{-0.8em}
\begin{itemize}
    \item \textbf{Separate objectives for code unlearning.}
    We distinguish recognizing target-related representations from suppressing target predictions, constructing a Fisher detector and a gradient-guided output direction from separate criteria.

    \item \textbf{The UNBIND framework.}
    UNBIND uses a calibrated gate to selectively update hidden states at a Transformer layer, keeping model and adapter weights fixed. Its inference-time module requires only two vectors and a few scalars.

    \item \textbf{Improved forgetting with limited utility loss.}
    Across four settings, UNBIND achieves the highest joint forgetting and utility score, exceeding the strongest evaluated baseline by up to \textbf{22.94 points} on a 100-point scale. It reduces F-BLEU by \textbf{97.27\% to 99.13\%} relative to the original models, with HumanEval+ and MBPP+ pass@1 drops of at most \textbf{1.22} and \textbf{1.59} percentage points.
\end{itemize}
\vspace{-1.0em}
\section{Related Work}
\vspace{-0.5em}
\paragraph{Code models and code unlearning.}
Trained on large code corpora, CodeLlama, StarCoder, and Qwen2.5-Coder support code completion, generation, and understanding~\citep{roziere2023codellama,li2023starcoder,hui2024qwen25coder}, but can also reproduce training implementations. Code unlearning seeks to suppress selected code while preserving programming ability, motivated by copyright concerns, insecure implementations, and deprecated APIs~\citep{jiang2026prod}.
\vspace{-1.0em}
\paragraph{Post-training unlearning.}
Post-training methods update model weights. Gradient ascent suppresses forget data, while gradient difference and retain KL regularization protect utility~\citep{maini2024tofu}. Preference-based approaches build on DPO~\citep{rafailov2024dpov3}, with NPO and SimNPO suppressing forget responses through negative preferences~\citep{zhang2024npo,fan2025simnpo}. FLAT learns template responses while suppressing forget outputs~\citep{wang2025flat}, ASU matches a teacher with smoothed attention~\citep{zarezade2026asu}, and RMU redirects forget representations while preserving retain states~\citep{li2024wmdpv7}. For code, CodeEraser combines gradient ascent on sensitive segments with gradient descent on surrounding code and KL constraints, targeting sensitive memorization rather than entire implementations~\citep{chu2026codeeraser}.
\vspace{-1.0em}
\paragraph{Representation intervention and inference-time unlearning.}
LLM representations support feature extraction and generation control, including truthfulness and refusal~\citep{zou2023representationengineering,turner2023activationaddition,shan2026sage,rimsky2024caa,li2023iti,arditi2024refusaldirection}. Conformal iteratively verifies and revises responses using verifier feedback, with conformal prediction calibrating the iteration budget~\citep{chowdhury2026conformal}. Divergence Decoding adjusts logits using the logit difference between two smaller auxiliary models, approximating retraining without forget data~\citep{merchant2026divergence}. GSS intervenes directly on hidden states, detecting memorization-related activations and selectively steering them~\citep{zhang2026gssgatedsubspacesteering}. Concurrent work includes ST$^2$U, which carries correction history across tokens to limit re-entry into restricted knowledge regions~\citep{chen2026st2u}. UNBIND applies selective hidden-state intervention to code unlearning, using a Fisher detector and a separately constructed gradient-guided output direction.
\vspace{-0.5em}
\section{Problem Setup}
\vspace{-0.75em}
We consider an autoregressive code model $M_\theta$ with model and adapter parameters $\theta$, a designated forget set $D_f$, and retain data $D_r$. The goal is to reduce reproduction of the target code while maintaining useful predictions and generation on retain data and independent programming tasks. We keep $\theta$ fixed and attach an intervention to one Transformer layer. Unlearning here denotes the resulting behavior while the intervention is active.

For a chosen decoder block, let $h_t\in\mathbb{R}^d$ denote its output at token position $t$ after its final residual addition, which contributes to predicting $x_{t+1}$. During construction, teacher forcing provides the target loss and its local gradient:
\vspace{-0.8em}
\begin{equation}
L_t=-\log p_\theta(x_{t+1}\mid x_{\leq t}),
\qquad g_t=\frac{\partial L_t}{\partial h_t}.
\label{eq:target-gradient}
\end{equation}
\vspace{-1em}

Here $x_{\leq t}$ is the input token prefix, $p_\theta$ is the conditional next-token probability, $L_t$ is the target-token negative log-likelihood, and $g_t$ is its gradient with respect to $h_t$.
We collect paired forget states and gradients as $H_f,G_f\in\mathbb{R}^{n_f\times d}$, and retain states as $H_r\in\mathbb{R}^{n_r\times d}$. The row counts $n_f$ and $n_r$ are the numbers of selected next-token prediction positions from forget and retain data, not source files. Separate calibration states drawn from training data determine the trigger threshold. Our learned model $M_{\mathrm{full}}$ and retain-only reference $M_{\mathrm{retrain}}$ share the pretrained model and low-rank adaptation (LoRA) initialization~\citep{hu2022lora}. The latter omits forget examples during adaptation and reproduction and utility measurements show how the intervention changes model behavior.
\vspace{-0.75em}
\begin{figure}[h]
\centering
\includegraphics[width=\linewidth]{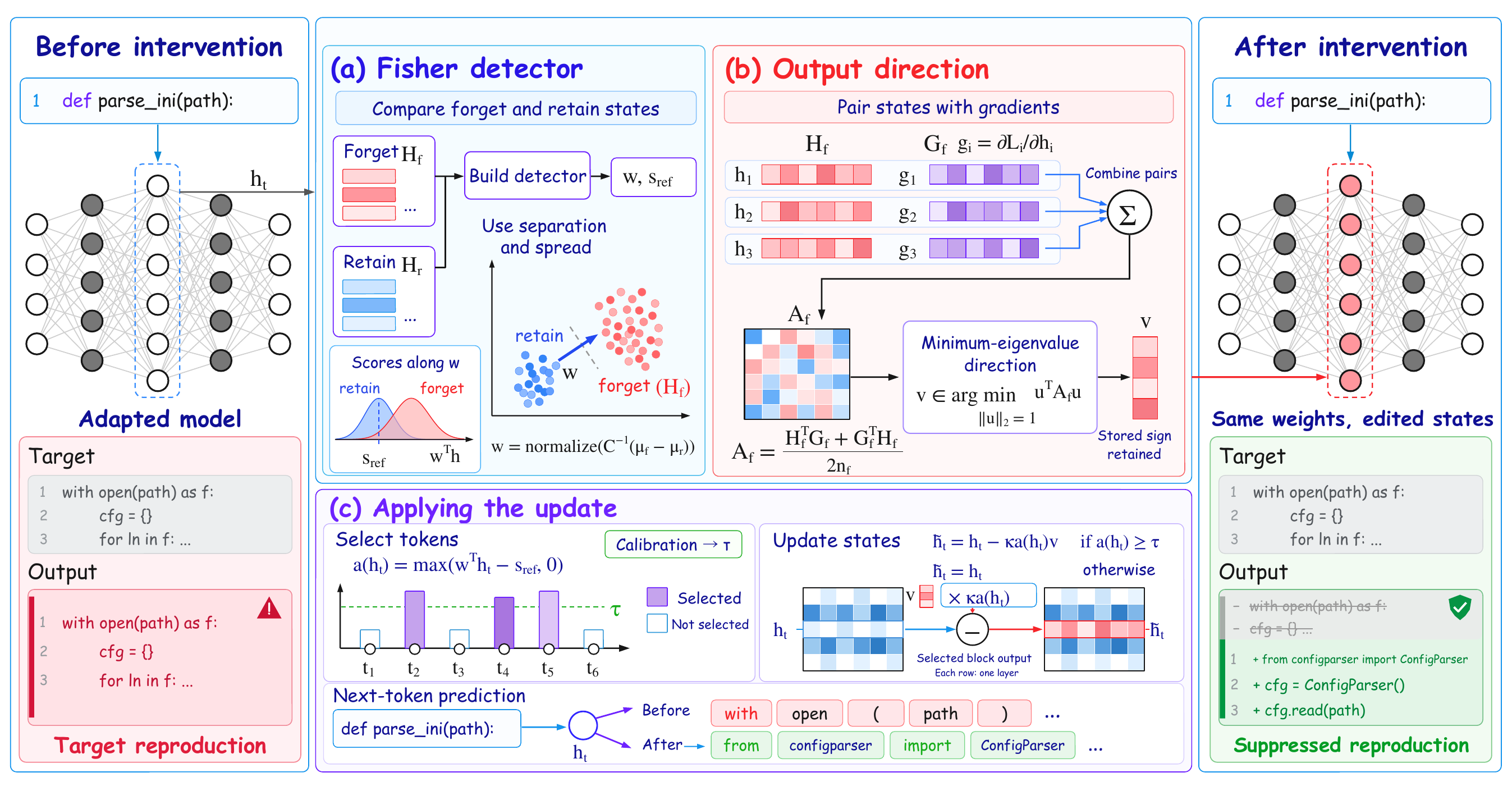}
\caption{\textbf{Framework of UNBIND.}
UNBIND separately constructs a Fisher detector to identify target-related states and an output direction to suppress target predictions. A calibrated gate then applies the intervention selectively at inference time while keeping model weights fixed.}
\label{fig:method}
\end{figure}
\vspace{-2em}
\section{UNBIND: Selective Representation Intervention}
\vspace{-0.75em}
Code unlearning is challenging because target implementations share representations with other programs, so suppressing them may harm broader programming ability. To address this issue, UNBIND separates which states to modify from how: a Fisher detector $w$ identifies target-related states, a separately constructed output direction $v$ specifies the update, and a calibrated gate controls when it is applied, as shown in Figure~\ref{fig:method}. 
\vspace{-0.75em}
\subsection{Detecting target-related states}
\vspace{-0.5em}
To suppress target code selectively, we need to identify the positions where an intervention should be applied during generation. Hidden states represent the context the model uses to predict the next token. Since our intervention directly modifies these representations, they also provide a natural basis for determining whether a position is associated with target code. We therefore collect hidden states from forget and retain code at the selected layer and compare the two groups to construct a detector.

A straightforward approach is to use the difference between the average hidden states of the two groups as the detector direction. However, a difference in the group averages does not necessarily mean that individual states can be reliably distinguished. If the states vary widely along this direction, some retain states may receive scores similar to those of forget states. As a result, using these scores to select intervention positions may also affect code that should be preserved. We therefore use a regularized Fisher discriminant, which measures the separation between the group means relative to the variation within each group. By reducing the influence of directions with large variation, the resulting detector captures differences that more reliably distinguish forget states from retain states.

For $c\in\{f,r\}$, let $n_c$ denote the number of selected token positions in the forget or retain group.
We write row $i$ of $H_c$ as a column vector $h_i^{(c)}\in\mathbb{R}^d$, where $d$ is the hidden dimension.
Giving each selected position equal weight within its group, we compute the mean $\mu_c$ and unbiased sample covariance $\Sigma_c$ as
\vspace{-0.5em}
\begin{equation}
\mu_c=\frac{1}{n_c}\sum_{i=1}^{n_c}h_i^{(c)},
\qquad
\Sigma_c=\frac{1}{n_c-1}\sum_{i=1}^{n_c}
(h_i^{(c)}-\mu_c)(h_i^{(c)}-\mu_c)^\top.
\label{eq:sample-statistics}
\end{equation}
\vspace{-1em}

The sum $\Sigma_f+\Sigma_r$ captures variation around the group means.
We use this matrix to adjust the mean difference $\mu_f-\mu_r$, reducing the influence of directions along which the states vary widely.
To stabilize the computation, we add a ridge term to form $C$ before computing the adjusted direction $\widetilde w$ and normalizing it to obtain the unit detector direction $w$:
\begin{align}
C &= \Sigma_f+\Sigma_r+\lambda\frac{\operatorname{tr}(\Sigma_f+\Sigma_r)}{d}I,
\quad \lambda=0.01,
\label{eq:covariance}\\
\widetilde w &= C^{-1}(\mu_f-\mu_r),
\qquad
w=\frac{\widetilde w}{\|\widetilde w\|_2}.
\label{eq:fisher}
\end{align}
\vspace{-1em}

Here, $I$ is the $d\times d$ identity matrix and $\operatorname{tr}$ denotes the trace.
The factor $\operatorname{tr}(\Sigma_f+\Sigma_r)/d$ scales the regularization strength $\lambda$ by the average variance, making the ridge term relative to the variation in the observed states.  

For a hidden state $h$, the detector score is $w^\top h$.
To express this score relative to retained code, we use the mean retain score as a reference $s_{\mathrm{ref}}$.
We then define the activation $a(h)$ as the positive excess above this reference:
\begin{equation}
s_{\mathrm{ref}}=\frac{1}{n_r}\sum_{i=1}^{n_r}w^\top h_i^{(r)}
=w^\top\mu_r,
\qquad
a(h)=\max(w^\top h-s_{\mathrm{ref}},0).
\label{eq:activation}
\end{equation}
\vspace{-1em}

The activation is zero at or below the retain reference and measures how far the score exceeds it otherwise.
A positive activation alone does not require an intervention.
The calibrated gate uses $a(h)$ to determine which positions should be modified.
\vspace{-0.75em}
\subsection{Constructing the output direction}
\vspace{-0.5em}
After identifying target-related states, we need to decide how to modify them to suppress target predictions. The detector direction may seem like a natural candidate for this update. However, this direction reflects differences between group means and variation within each group, not how changes in a state affect target-token loss. Moving against it can therefore lower the detector score without making the target token less likely. We instead use gradients of the target-token loss to construct a separate output direction. These gradients describe how the loss responds locally to changes in the hidden state, allowing us to compare directions by their predicted effect on the loss.

To compare candidate directions, consider the auxiliary update $\delta h=-\eta(u^\top h)u$, where $h$ is a hidden state, $u$ is a unit direction, and $\eta\geq0$ is the step size. The scalar $u^\top h$ is the projection of the state onto $u$, so this update moves against that component. For each of the $n_f$ positions in the forget data used for fitting, let $h_i^{(f)}$ be its hidden state and $g_i^{(f)}$ the gradient of its target-token loss with respect to that state. When one state is perturbed with the others held fixed, the local loss change is approximated by the inner product of its gradient and the perturbation. Substituting the auxiliary update and averaging over positions gives
\vspace{-0.5em}
\begin{equation}
\mathbb{E}_f[\delta L]\approx-\frac{\eta}{n_f}
\sum_{i=1}^{n_f}(u^\top h_i^{(f)})\bigl((g_i^{(f)})^\top u\bigr).
\label{eq:proxy}
\end{equation}
\vspace{-1em}

Here $\delta L$ denotes the local target-token loss change, and $\mathbb{E}_f$ denotes the empirical average over the $n_f$ positions. Because of the negative sign, maximizing the predicted loss increase for $\eta>0$ requires minimizing the average product in Equation~\ref{eq:proxy}.

To express this objective in matrix form, let $H_f$ and $G_f$ contain the paired states and gradients as corresponding rows, without centering either group of vectors. The average product is then $u^\top H_f^\top G_f u/n_f$. Replacing $H_f^\top G_f/n_f$ with its symmetric part leaves this scalar unchanged. We denote the resulting symmetric matrix by $A_f$ and select the unit direction $v$ that minimizes its quadratic form:
\begin{equation}
A_f=\frac{H_f^\top G_f+G_f^\top H_f}{2n_f},
\qquad
v\in\underset{\|u\|_2=1}{\arg\min}\ u^\top A_f u.
\label{eq:direction}
\end{equation}
The solution $v$ is a unit eigenvector associated with the smallest eigenvalue $\lambda_{\min}(A_f)$. Substituting this direction into Equation~\ref{eq:proxy} gives the maximum auxiliary first-order loss change $-\eta\lambda_{\min}(A_f)$, which is positive for $\eta>0$ only when $\lambda_{\min}(A_f)<0$. Using only forget states and their gradients, this criterion selects the output direction according to the local loss response to the auxiliary update, rather than separation between forget and retain states.
\vspace{-0.75em}
\subsection{Calibrating and applying the update}
\vspace{-0.5em}
\begingroup
\let\mergedlabel\label

A positive detector activation suggests that a state is related to
target code, but modifying it may also affect useful predictions.
A low threshold can admit changes that harm utility.
Raising the threshold can reduce this interference but may leave
target reproduction insufficiently suppressed.
Even with a selective gate, an overly strong update can harm utility.
We therefore calibrate both the threshold and intervention strength
under fixed retain budgets, using data separate from those used to
construct the detector and output axis.

Let $\mathcal A_{\mathrm{cal}}^+$ contain the positive detector
activations from unmodified forget calibration training states.
A prespecified finite set $\mathcal Q\subseteq[0,1]$ gives the candidate
quantile levels. For each $q$, we compute a threshold $\tau_q$ by
linear interpolation with repeated values retained.
The threshold defines a gate $m$ that admits a hidden state $h$
when its activation reaches $\tau_q$:
\begin{equation}
\tau_q =
\operatorname{Quantile}_q\!\left(\mathcal A_{\mathrm{cal}}^+\right),
\quad q\in\mathcal Q,
\qquad
m(h;\tau)=\mathbf{1}[a(h)\geq\tau].
\label{eq:threshold}
\end{equation}

The quadratic objective in Section~4.2 gives a unit eigenvector
$v_0$ but does not distinguish its two signs.
We choose a sign for each gate based on the states that it admits.
At the $n_c$ unmodified forget calibration training positions,
write $a_i=a(h_i)$.
Let $g_i=\nabla_{h_i}\ell_i^{\mathrm{NLL}}$ be the gradient of the
target token NLL $\ell_i^{\mathrm{NLL}}$ with respect to its
prediction state $h_i$.
The gated activation $b_{i,q}$ weights each gradient projection
onto $v_0$, giving the weighted average $c_q$.
Because the update subtracts the chosen direction, we orient
$v_q$ so that the weighted average projection onto it is nonpositive:
\begin{equation}
b_{i,q}=a_i\mathbf{1}[a_i\geq\tau_q],
\qquad
c_q=
\frac{\sum_{i=1}^{n_c}b_{i,q}\,g_i^\top v_0}
{\sum_{i=1}^{n_c}b_{i,q}},
\qquad
v_q=
\begin{cases}
v_0, & c_q<0,\\
-v_0, & c_q>0.
\end{cases}
\label{eq:calibration-response}
\mergedlabel{eq:direction-sign}
\end{equation}
When $c_q=0$, we choose the sign that makes the component
with the largest absolute value positive.

For each candidate gate, set $\tau=\tau_q$ and $v=v_q$.
At position $t$, we modify the hidden state $h_t$ with
intervention strength $\kappa\geq0$:
\begin{equation}
\widetilde h_t
=
h_t-\kappa\,m(h_t;\tau)\,a(h_t)\,v.
\label{eq:update}
\end{equation}
A state outside the gate remains unchanged.
For an admitted state, the update has magnitude $\kappa a(h_t)$
because $\|v\|_2=1$.
The chosen sign preserves the quadratic objective and maximizes
the predicted average first-order NLL increase over the two
orientations.
This estimate treats each unmodified state as perturbed separately
and does not guarantee an increase in the loss of a full sequence.
Changing $\kappa$ only scales this local estimate, so the sign
need not be recalibrated.

For each $q$, we fix $(\tau_q,v_q)$ and scan strengths from a
prespecified finite set $\mathcal K\subseteq[0,\infty)$ on separate
calibration validation data.
We use teacher forcing to measure forget NLL $L_f^{\mathrm{val}}$
and the retain NLL increase $\Delta L_r$ relative to the original model.
We also compute retain KL $D_r$ from the original predictive
distributions to the intervened ones.
Each measure averages over all target tokens in its subset.
We first require $D_r\leq\epsilon_{\mathrm{KL}}$ and
$\Delta L_r\leq\epsilon_{\mathrm{NLL}}$, where the fixed budgets
$\epsilon_{\mathrm{KL}}$ and $\epsilon_{\mathrm{NLL}}$ bound the
allowed divergence and NLL increase, respectively.
Candidates that pass these checks are then evaluated by free
generation against a prespecified retain BLEU budget.
Let $\mathcal F\subseteq\mathcal Q\times\mathcal K$ contain the
candidates satisfying all three constraints.
If $\mathcal F\neq\varnothing$, we select the candidate with the
highest forget validation NLL:
\begin{equation}
(q^\star,\kappa^\star)
\in
\arg\max_{(q,\kappa)\in\mathcal F}
L_f^{\mathrm{val}}(q,\kappa).
\label{eq:calibration-selection}
\end{equation}
\vspace{-1em}

The selected $\tau=\tau_{q^\star}$, $v=v_{q^\star}$, and
$\kappa=\kappa^\star$ remain fixed during prompt processing and
cached decoding.
Deployment stores $(w,s_{\mathrm{ref}},v,\tau,\kappa)$ and leaves
model and adapter weights unchanged.
No gradients are needed during inference, and the update adds
$O(d)$ operations per position, where $d$ is the hidden dimension.
Retain data inform the detector and reference score as well as
the calibration constraints.
Section~5 evaluates utility preservation.
Appendix~\ref{app:construction:calibration} provides calibration
details, while Appendix~\ref{app:construction:inference} derives
the update magnitude and local loss response.

\endgroup
\vspace{-1em}
\section{Experiments}
\vspace{-0.75em}
We evaluate target code reproduction and code utility, then test access under changed prompts and sampling, intervention strength, components, languages, and related code.
\vspace{-1em}
\subsection{Experimental protocol}
\vspace{-0.5em}
\paragraph{Models and data.}
We use Qwen2.5-Coder-7B and CodeLlama-7B~\citep{hui2024qwen25coder,roziere2023codellama} on CodeSearchNet (CSN) and The Stack~\citep{husain2019codesearchnet,kocetkov2022stack}, forming QwenCSN, QwenStack, CLCSN, and CLStack. Each corpus contains 300 Python forget-training and 8,280 retain-training examples. Learned and retain-only models use the same LoRA initialization and training recipe, except that retain-only training excludes forget slots. Appendices~\ref{app:protocol:data} and~\ref{app:protocol:models}
provide data construction and model adaptation details.
\vspace{-1em}
\paragraph{Intervention and baselines.}
UNBIND uses $\kappa=20$ and constructs the direction using only forget states and gradients. We scan all 28 Qwen layers and 32 CodeLlama layers, selecting layers 23, 24, 29, and 28 for the four settings. We compare against fourteen baselines: GA, GradDiff, NPO-KL, SimNPO-KL, DPO, FLAT, PROD, GSS, RMU, ASU-KL, Task Vector, CodeEraser, Divergence Decoding (DD), and Conformal Unlearning (CU). Layer configurations are described in
Appendix~\ref{app:protocol:models}. UNBIND construction and
calibration and baseline implementations are detailed in
Appendices~\ref{app:construction} and~\ref{app:baselines}.
\vspace{-1em}
\paragraph{Metrics.}
F-BLEU measures lexical reproduction of forget-set continuations, with lower values indicating better forgetting~\citep{papineni2002bleu}. Utility metrics include retain perplexity (PPL) from mean token NLL, retain BLEU (R-BLEU) on examples with valid tokenizer continuations, and HumanEval+ (HE+)/MBPP+ counts of problems passing both base and extended tests~\citep{chen2021humaneval,austin2021mbpp,liu2023evalplus}. Generation uses one greedy completion per example, BF16, a 128-token cap, and batch size four. The Forgetting--Utility Harmonic Score (FU-H) combines forgetting and utility:
\vspace{-1em}
\begin{equation}
\mathrm{FR}=\max\!\left(0,1-\frac{F}{F_{\mathrm{full}}}\right),\quad
\mathrm{UR}=\left[\prod_{j=1}^{4}\min(1,r_j)\right]^{1/4},\quad
\mathrm{FU\mbox{-}H}=100\frac{2\mathrm{FR}\,\mathrm{UR}}{\mathrm{FR}+\mathrm{UR}}.
\label{eq:joint-score}
\end{equation}
Here $F$ is F-BLEU, ``full'' denotes the original model, and $r_j$ are
$\mathrm{PPL}_{\mathrm{full}}/\mathrm{PPL}$,
$\mathrm{R\mbox{-}BLEU}/\mathrm{R\mbox{-}BLEU}_{\mathrm{full}}$,
$\mathrm{HE+}/\mathrm{HE+}_{\mathrm{full}}$, and
$\mathrm{MBPP+}/\mathrm{MBPP+}_{\mathrm{full}}$.
Capping ratios at one prevents utility gains from offsetting losses.
We set FU-H to zero when $\mathrm{FR}=\mathrm{UR}=0$.
See Appendix~\ref{app:protocol:metrics} for evaluation details.

\subsection{Suppression and utility across models}
\vspace{-0.5em}
Tables~\ref{tab:main_qwen} and~\ref{tab:main_codellama} summarize the results. Across all four settings, UNBIND reduces F-BLEU by 97.3--99.1\% and achieves the highest FU-H, ranging from 98.57 to 98.78. Relative to $M_{\mathrm{full}}$, retain PPL increases by at most 1.6\%, R-BLEU changes by at most 0.0029, and HE+ and MBPP+ pass counts differ by at most two and six. Low F-BLEU alone, however, does not imply selective forgetting: on CL-CSN, GSS and GradDiff achieve even lower F-BLEU than UNBIND, but both reduce HE+ and MBPP+ to zero. On QwenStack, UNBIND reaches 0.0042 F-BLEU versus 0.0062 for GSS while retaining 91/236 HE+/MBPP+ passes versus 58/15.
\vspace{-1.75em}
\begin{table}[!htbp]
\centering
\caption{Qwen results. HE+/MBPP+: pass counts out of 164/378. Colors mark \colorbox{rankfirst}{\textbf{first}}, \colorbox{ranksecond}{\underline{second}}, and \colorbox{rankthird}{\textit{third}} distinct values; raw metrics use displayed precision, FU-H unrounded scores.}
\label{tab:main_qwen}
\fontsize{7.6}{9.1}\selectfont
\setlength{\tabcolsep}{1.35pt}
\renewcommand{\arraystretch}{1.02}
\begin{tabular*}{\linewidth}{@{\extracolsep{\fill}}lrrrrrrrrrrrr@{}}
\toprule
 & \multicolumn{6}{c}{\textbf{CodeSearchNet}} & \multicolumn{6}{c}{\textbf{The Stack}} \\
\cmidrule(lr){2-7}\cmidrule(l){8-13}
Method & F-BLEU$\!\downarrow$ & R-PPL$\!\downarrow$ & R-BLEU$\!\uparrow$ & HE+$\!\uparrow$ & MBPP+$\!\uparrow$ & \textbf{FU-H}$\!\uparrow$ & F-BLEU$\!\downarrow$ & R-PPL$\!\downarrow$ & R-BLEU$\!\uparrow$ & HE+$\!\uparrow$ & MBPP+$\!\uparrow$ & \textbf{FU-H}$\!\uparrow$ \\
\midrule
$M_{\mathrm{full}}$ & 0.4847 & 3.0046 & 0.3082 & 78 & 225 & 0.00 & 0.1531 & 2.0861 & 0.3874 & 91 & 233 & 0.00 \\
$M_{\mathrm{retrain}}$ & 0.0907 & 2.9724 & 0.3060 & 71 & 227 & 88.66 & 0.0872 & 2.0818 & 0.3907 & 87 & 233 & 59.99 \\
$M_{\mathrm{base}}$ & 0.0847 & 4.1574 & 0.2543 & 121 & 246 & 85.12 & 0.0850 & 2.1894 & 0.3702 & 121 & 246 & 61.13 \\
\midrule
GA & 0.3031 & 2.9654 & \cellcolor{rankthird}\textit{0.3100} & \cellcolor{rankthird}\textit{79} & 225 & 54.52 & 0.0754 & 2.1049 & 0.3762 & 89 & \cellcolor{rankfirst}\textbf{254} & 67.02 \\
GradDiff & \cellcolor{rankfirst}\textbf{0.0041} & 26209.519 & 0.0676 & 0 & 0 & 0.00 & 0.0613 & 2.1513 & 0.3644 & \cellcolor{rankthird}\textit{91} & 242 & \cellcolor{rankthird}\textit{74.35} \\
NPO-KL & 0.2135 & \cellcolor{ranksecond}\underline{2.9600} & 0.3025 & \cellcolor{rankthird}\textit{79} & \cellcolor{ranksecond}\underline{228} & 71.63 & 0.1274 & 2.0771 & 0.3856 & 88 & 239 & 28.76 \\
SimNPO-KL & 0.0984 & 2.9833 & 0.3005 & \cellcolor{rankthird}\textit{79} & 220 & \cellcolor{ranksecond}\underline{88.23} & 0.0979 & \cellcolor{rankthird}\textit{2.0768} & 0.3838 & 89 & \cellcolor{ranksecond}\underline{248} & 52.89 \\
DPO & 0.2447 & \cellcolor{rankthird}\textit{2.9604} & \cellcolor{rankfirst}\textbf{0.3104} & \cellcolor{rankfirst}\textbf{83} & 225 & 66.23 & 0.1361 & 2.0786 & 0.3859 & \cellcolor{rankfirst}\textbf{93} & \cellcolor{rankthird}\textit{243} & 20.05 \\
FLAT & 0.0761 & 3.2647 & 0.2795 & 43 & 200 & \cellcolor{rankthird}\textit{82.08} & 0.0877 & 2.1191 & 0.3663 & 86 & 227 & 59.17 \\
PROD & 0.2302 & 3.0215 & 0.3048 & 64 & 212 & 67.22 & 0.1374 & 2.0864 & \cellcolor{rankthird}\textit{0.3863} & 88 & 233 & 18.60 \\
Task Vector & 0.2495 & 2.9929 & 0.2944 & 74 & \cellcolor{rankfirst}\textbf{231} & 64.81 & 0.1284 & \cellcolor{ranksecond}\underline{2.0754} & 0.3828 & 86 & 241 & 27.73 \\
RMU & 0.4810 & 3.0063 & 0.3067 & 76 & 221 & 1.50 & 0.1514 & 2.0862 & \cellcolor{rankfirst}\textbf{0.3886} & 89 & 231 & 2.26 \\
ASU-KL & 0.2029 & 3.0790 & 0.2925 & 69 & 215 & 71.87 & 0.1075 & 2.1074 & 0.3862 & 86 & 232 & 45.72 \\
GSS & 0.0920 & 3.5534 & 0.2366 & 72 & 139 & 79.48 & \cellcolor{ranksecond}\underline{0.0062} & 6.2591 & 0.2689 & 58 & 15 & 47.11 \\
DD & 0.0883 & 3.5559 & 0.2568 & 45 & 167 & 77.75 & \cellcolor{rankthird}\textit{0.0385} & 2.4478 & 0.3202 & 68 & 151 & \cellcolor{ranksecond}\underline{75.63} \\
CodeEraser & 0.3872 & 2.9649 & 0.3071 & \cellcolor{ranksecond}\underline{82} & \cellcolor{rankthird}\textit{227} & 33.49 & 0.1494 & 2.0825 & 0.3855 & \cellcolor{ranksecond}\underline{92} & 236 & 4.72 \\
CU & \cellcolor{rankthird}\textit{0.0378} & \cellcolor{rankfirst}\textbf{2.2191} & 0.2352 & 69 & 53 & 74.96 & 0.0547 & \cellcolor{rankfirst}\textbf{1.8824} & 0.3413 & 80 & 73 & 67.10 \\
\midrule
\textbf{UNBIND} & \cellcolor{ranksecond}\underline{0.0042} & 3.0345 & \cellcolor{ranksecond}\underline{0.3101} & 76 & 219 & \cellcolor{rankfirst}\textbf{98.78} & \cellcolor{rankfirst}\textbf{0.0042} & 2.0943 & \cellcolor{ranksecond}\underline{0.3883} & \cellcolor{rankthird}\textit{91} & 236 & \cellcolor{rankfirst}\textbf{98.57} \\
\bottomrule
\end{tabular*}
\end{table}
\vspace{-1em}

The reference models clarify the effect of adaptation. F-BLEU for $M_{\mathrm{full}}$ is 1.8--5.7 times that of $M_{\mathrm{base}}$, whereas $M_{\mathrm{retrain}}$ remains close to the base level. UNBIND achieves 94.4--95.8\% lower F-BLEU than $M_{\mathrm{base}}$ and lower F-BLEU than $M_{\mathrm{retrain}}$ in every setting, suppressing reproduction beyond returning to either reference level. It also maintains lower retain PPL and higher retain BLEU than $M_{\mathrm{base}}$ in three settings. These comparisons establish suppression below the pre-adaptation reproduction level, but base-model F-BLEU alone does not establish that the targets were memorized during pretraining.

\vspace{-1.25em}
\begin{table}[!htbp]
\centering
\caption{CodeLlama results with selected baselines. Metrics and ranking conventions follow Table~\ref{tab:main_qwen}; rankings include only the displayed non-reference methods. Complete results appear in Table~\ref{tab:full_codellama}.}
\label{tab:main_codellama}
\fontsize{7.6}{9.1}\selectfont
\setlength{\tabcolsep}{1.35pt}
\renewcommand{\arraystretch}{1.02}
\begin{tabular*}{\linewidth}{@{\extracolsep{\fill}}lrrrrrrrrrrrr@{}}
\toprule
 & \multicolumn{6}{c}{\textbf{CodeSearchNet}} & \multicolumn{6}{c}{\textbf{The Stack}} \\
\cmidrule(lr){2-7}\cmidrule(l){8-13}
Method & F-BLEU$\!\downarrow$ & R-PPL$\!\downarrow$ & R-BLEU$\!\uparrow$ & HE+$\!\uparrow$ & MBPP+$\!\uparrow$ & \textbf{FU-H}$\!\uparrow$ & F-BLEU$\!\downarrow$ & R-PPL$\!\downarrow$ & R-BLEU$\!\uparrow$ & HE+$\!\uparrow$ & MBPP+$\!\uparrow$ & \textbf{FU-H}$\!\uparrow$ \\
\midrule
$M_{\mathrm{full}}$ & 0.4600 & 2.2521 & 0.3949 & 38 & 125 & 0.00 & 0.1434 & 1.7330 & 0.4736 & 36 & 163 & 0.00 \\
$M_{\mathrm{retrain}}$ & 0.0894 & 2.2526 & 0.3940 & 40 & 114 & 88.30 & 0.0685 & 1.7320 & 0.4754 & 36 & 162 & 68.56 \\
$M_{\mathrm{base}}$ & 0.0963 & 2.6220 & 0.3474 & 41 & 158 & 85.57 & 0.0651 & 1.7160 & 0.4732 & 41 & 158 & 70.45 \\
\midrule
SimNPO-KL & 0.1330 & \cellcolor{ranksecond}\underline{2.2393} & 0.3802 & \cellcolor{rankthird}\textit{33} & 77 & \cellcolor{ranksecond}\underline{77.31} & 0.0834 & \cellcolor{rankfirst}\textbf{1.7224} & \cellcolor{ranksecond}\underline{0.4817} & \cellcolor{rankfirst}\textbf{40} & \cellcolor{rankthird}\textit{161} & 58.91 \\
FLAT & \cellcolor{rankthird}\textit{0.0301} & 2.4337 & 0.3151 & 16 & 46 & 71.70 & \cellcolor{ranksecond}\underline{0.0036} & 1.7331 & 0.4769 & 35 & 154 & \cellcolor{ranksecond}\underline{97.68} \\
PROD & 0.3073 & \cellcolor{rankfirst}\textbf{2.2372} & \cellcolor{rankthird}\textit{0.3907} & 30 & \cellcolor{rankthird}\textit{107} & 48.56 & 0.1244 & \cellcolor{rankthird}\textit{1.7270} & 0.4752 & \cellcolor{ranksecond}\underline{38} & 154 & 23.31 \\
Task Vector & 0.2547 & \cellcolor{rankthird}\textit{2.2428} & \cellcolor{ranksecond}\underline{0.3937} & \cellcolor{rankfirst}\textbf{39} & \cellcolor{ranksecond}\underline{108} & 61.01 & 0.1045 & \cellcolor{ranksecond}\underline{1.7230} & \cellcolor{rankthird}\textit{0.4772} & \cellcolor{ranksecond}\underline{38} & \cellcolor{ranksecond}\underline{162} & 42.65 \\
GSS & \cellcolor{rankfirst}\textbf{0.000014} & 624186.8588 & 0.0003 & 0 & 0 & 0.00 & 0.0994 & 1.7737 & \cellcolor{rankfirst}\textbf{0.4830} & 33 & \cellcolor{ranksecond}\underline{162} & 46.66 \\
DD & 0.1143 & 2.6876 & 0.3452 & 24 & 63 & \cellcolor{rankthird}\textit{72.21} & \cellcolor{rankthird}\textit{0.0560} & 2.0080 & 0.4181 & 24 & 100 & \cellcolor{rankthird}\textit{67.10} \\
\midrule
\textbf{UNBIND} & \cellcolor{ranksecond}\underline{0.0054} & 2.2588 & \cellcolor{rankfirst}\textbf{0.3978} & \cellcolor{ranksecond}\underline{36} & \cellcolor{rankfirst}\textbf{129} & \cellcolor{rankfirst}\textbf{98.71} & \cellcolor{rankfirst}\textbf{0.0027} & 1.7599 & 0.4707 & \cellcolor{rankthird}\textit{36} & \cellcolor{rankfirst}\textbf{163} & \cellcolor{rankfirst}\textbf{98.77} \\
\bottomrule
\end{tabular*}
\end{table}
\vspace{-1em}

\subsection{Strength-dependent tradeoffs}
\begin{figure}[!htp]
\centering
\includegraphics[width=1\linewidth]{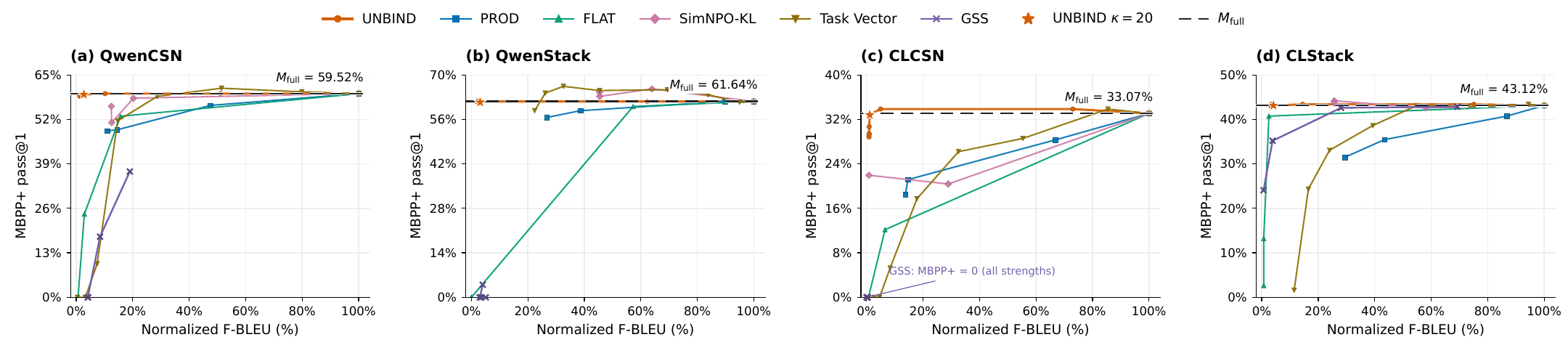}
\caption{MBPP+ pass@1 (\%, denominator 378) versus normalized F-BLEU ($100F/F_{\mathrm{full}}$). Dashed lines mark original utility; stars mark UNBIND at $\kappa=20$ with the chosen layer configurations. }
\label{fig:strength}
\end{figure}

\vspace{-0.5em}
We evaluate UNBIND and the baselines at different strengths across all four settings, with Figure~\ref{fig:strength} showing the resulting tradeoffs between target reproduction and MBPP+. On three settings, UNBIND keeps MBPP+ near its original level as F-BLEU decreases. In contrast, several baselines achieve low reproduction only with substantial utility losses. These curves show that UNBIND provides a more favorable balance between forgetting and utility across multiple strengths.
\vspace{-2em}
\subsection{What the detector and output direction contribute}
\begin{figure}[!htb]
\centering
\includegraphics[width=\linewidth]{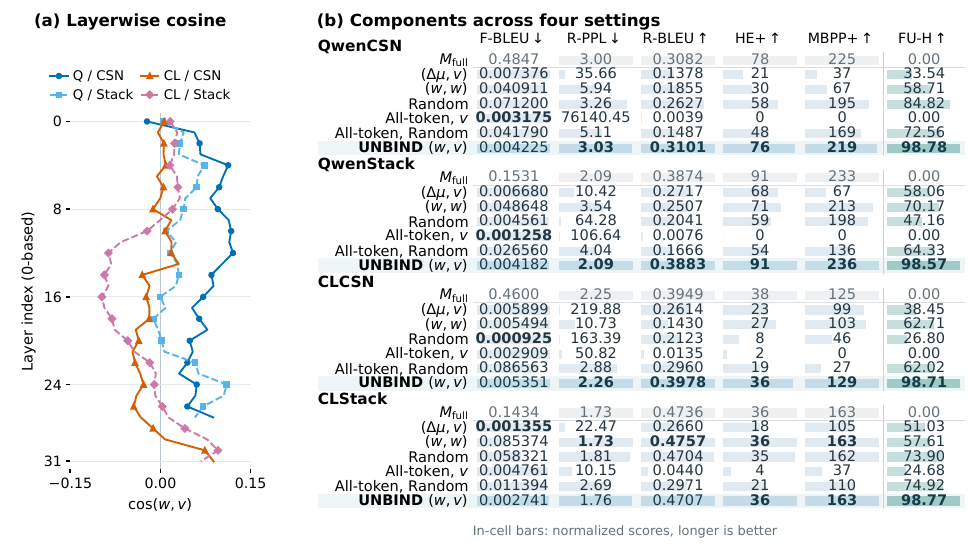}
\caption{(a) Layerwise $\cos(w,v)$; Q/CL denote Qwen/Code Llama. (b) Four-setting ablations; pairs denote detector/output direction. Cells show raw values; bars show normalized scores, with longer bars indicating better results (Equation~\ref{eq:joint-score}). Bold marks best intervention values, including ties. HE+/MBPP+ count passes out of 164/378.}
\label{fig:direction-ablation}
\end{figure}

\vspace{-0.5em}
Figure~\ref{fig:direction-ablation}(a) shows weak alignment between detector $w$ and output direction $v$ across layers, while Figure~\ref{fig:direction-ablation}(b) examines their respective roles.

\vspace{-0.5em}
Replacing the Fisher detector with mean difference lowers F-BLEU but raises retain PPL to 10.42--219.88, while reusing $w$ as the output direction yields higher F-BLEU than UNBIND in all four settings. Low reproduction alone is also insufficient: on CLCSN, the Random ablation reaches F-BLEU 0.000925 but raises retain PPL to 163.39. Applying $v$ at every position reduces MBPP+ to zero in three settings and 37 in CLStack, while all-token random updates achieve only 62.02--74.92 FU-H. In contrast, UNBIND achieves 98.57--98.78 FU-H while remaining within 2 HE+ and 6 MBPP+ passes of $M_{\mathrm{full}}$. These results support separating detection from suppression and applying the output update selectively with activation-dependent magnitudes.
\vspace{-0.75em}
\subsection{Target reproduction and multilingual performance}
\vspace{-0.5em}
\label{sec:access}
\label{sec:multilingual}

We evaluate target reproduction under six fixed conditions: the original prompt, 16-, 32-, or 64-token target-prefix injection, instruction rephrasing, and code fences. Each condition uses one greedy and four sampled completions, capped at 256 new tokens. We report the number of targets with $\geq50/\geq100$-token contiguous exact spans and the mean best recovery $\overline R$, defined as the maximum exact-span-to-withheld-suffix ratio across attempts, averaged over 300 targets. We also evaluate Python, Java, and JavaScript forgetting tasks
on both model families, with utility measured by retain PPL,
R-BLEU, and Python HE+/MBPP+. Detailed protocols and CodeLlama
results are in Appendices~\ref{app:access} and~\ref{app:multilingual}.
\vspace{-1em}
\begin{table}[!htbp]
\centering
\caption{Target reproduction and multilingual results for Qwen2.5-Coder-7B. Counts report targets with $\geq50/\geq100$-token exact spans out of 300; mean best recovery $\overline R$, FR, and UR are percentages. }
\label{tab:access-multilingual}
\fontsize{7.6}{9.1}\selectfont
\setlength{\tabcolsep}{1.1pt}
\renewcommand{\arraystretch}{1.0}
\begin{tabular*}{\linewidth}{@{\extracolsep{\fill}}l*{13}{r}@{}}
\toprule
 & \multicolumn{4}{c}{Target reproduction $\downarrow$}
 & \multicolumn{9}{c}{Multilingual: forget language} \\
\cmidrule(lr){2-5}\cmidrule(l){6-14}
 & \multicolumn{2}{c}{CSN}
 & \multicolumn{2}{c}{Stack}
 & \multicolumn{3}{c}{Python}
 & \multicolumn{3}{c}{Java}
 & \multicolumn{3}{c}{JavaScript} \\
\cmidrule(lr){2-3}\cmidrule(lr){4-5}
\cmidrule(lr){6-8}\cmidrule(lr){9-11}\cmidrule(l){12-14}
Method
 & Counts & $\overline R$
 & Counts & $\overline R$
 & FR$\uparrow$ & UR$\uparrow$ & FU-H$\uparrow$
 & FR$\uparrow$ & UR$\uparrow$ & FU-H$\uparrow$
 & FR$\uparrow$ & UR$\uparrow$ & FU-H$\uparrow$ \\
\midrule
$M_{\mathrm{full}}$
 & 201/65 & 79.71 & 188/93 & 33.46
 & 0.00 & 100.00 & 0.00
 & 0.00 & 100.00 & 0.00
 & 0.00 & 100.00 & 0.00 \\
$M_{\mathrm{retrain}}$
 & 11/1 & 22.46 & 50/15 & 7.96
 & 80.84 & 98.82 & 88.93
 & 62.11 & 97.76 & 75.96
 & 64.45 & 100.00 & 78.38 \\
\midrule
PROD
 & 63/13 & 60.00 & 144/67 & 16.27
 & 51.75 & 90.11 & 65.74
 & 76.30 & 90.87 & 82.95
 & 69.00 & 95.79 & 80.22 \\
GSS
 & 61/14 & 53.81 & 13/\textbf{0} & 3.31
 & 58.95 & 97.98 & 73.61
 & 41.73 & 91.13 & 57.25
 & 34.80 & 95.58 & 51.02 \\
SimNPO-KL
 & 7/2 & 25.00 & 77/27 & 10.89
 & 81.46 & 98.10 & 89.01
 & 99.15 & 47.65 & 64.37
 & 97.96 & 62.95 & 76.65 \\
FLAT
 & 9/3 & 22.68 & 59/23 & 8.49
 & 84.77 & 84.26 & 84.51
 & \textbf{99.33} & 5.75 & 10.87
 & \textbf{99.08} & 7.03 & 13.13 \\
Task Vector
 & 62/13 & 56.55 & 112/47 & 12.62
 & 48.63 & 97.19 & 64.82
 & 45.22 & 95.18 & 61.31
 & 40.77 & \textbf{99.70} & 57.88 \\
\rowcolor{oursrow}
\textbf{UNBIND}
 & \textbf{0}/\textbf{0} & \textbf{6.45}
 & \textbf{1}/\textbf{0} & \textbf{1.30}
 & \textbf{99.13} & \textbf{98.44} & \textbf{98.78}
 & 97.08 & \textbf{99.85} & \textbf{98.45}
 & 98.81 & 99.33 & \textbf{99.07} \\
\bottomrule
\end{tabular*}
\end{table}
\vspace{-1em}

Under this access budget, UNBIND leaves no $\geq100$-token matches and only 0/1 targets with $\geq50$-token matches on QwenCSN/QwenStack, with the lowest $\overline R$ of 6.45\%/1.30\%. Across both model families and all three languages, it maintains FR above 97\% and achieves the highest FU-H in all six settings, showing consistent forgetting with little utility loss.
\vspace{-0.75em}
\subsection{Retention on related code}
\vspace{-0.5em}
We measure mean source NLL on retain code in its original format across four groups: same-language neighbors, cross-language neighbors, format-matched code, and ordinary retain code. To identify related retain code, we use UniXcoder~\citep{guo2022unixcoder} to retrieve candidates similar to the forget code in embedding space. We then review these candidates for semantic relatedness, checking whether they implement similar functionality or computational logic. Group sizes are summarized in Appendix~\ref{app:protocol:data}.
\vspace{-1.5em}
\begin{table}[!htbp]
\centering
\caption{Related-code retention on Qwen2.5-Coder-7B. Cells report F-BLEU or mean source NLL; superscripts give $100(x/x_{\mathrm{full}}-1)$ (\%). Ranking follows Tables~\ref{tab:main_qwen}--\ref{tab:main_codellama}.}
\label{tab:neighbors}
\fontsize{7.2}{8.6}\selectfont
\setlength{\tabcolsep}{1.2pt}
\renewcommand{\arraystretch}{1.12}
\newcommand{\neighborchange}[1]{\textsuperscript{\fontsize{5}{5}\selectfont$#1$}}
\resizebox{\linewidth}{!}{%
\begin{tabular}{@{}lrrrrrrrrrr@{}}
\toprule
 & \multicolumn{5}{c}{\textbf{CodeSearchNet}} & \multicolumn{5}{c}{\textbf{The Stack}} \\
\cmidrule(lr){2-6}\cmidrule(l){7-11}
 & Forget & \multicolumn{4}{c}{Retained utility (source NLL)} & Forget & \multicolumn{4}{c}{Retained utility (source NLL)} \\
\cmidrule(lr){3-6}\cmidrule(lr){8-11}
Method & \shortstack{F-BLEU\\$\downarrow$} & \shortstack{Same lang.\\$\downarrow$} & \shortstack{Cross lang.\\$\downarrow$} & \shortstack{Format\\$\downarrow$} & \shortstack{Ordinary\\$\downarrow$} & \shortstack{F-BLEU\\$\downarrow$} & \shortstack{Same lang.\\$\downarrow$} & \shortstack{Cross lang.\\$\downarrow$} & \shortstack{Format\\$\downarrow$} & \shortstack{Ordinary\\$\downarrow$} \\
\midrule
$M_{\mathrm{full}}$
 & 0.484702\neighborchange{}
 & 0.7430\neighborchange{}
 & 0.6728\neighborchange{}
 & 0.8559\neighborchange{}
 & 0.6929\neighborchange{}
 & 0.153138\neighborchange{}
 & 0.9404\neighborchange{}
 & 0.8373\neighborchange{}
 & 0.7796\neighborchange{}
 & 0.7858\neighborchange{} \\
\midrule
FLAT
 & \cellcolor{ranksecond}\underline{0.076068}\neighborchange{\mathord{\downarrow}84.31}
 & 0.9634\neighborchange{\mathord{\uparrow}29.66}
 & 0.7844\neighborchange{\mathord{\uparrow}16.59}
 & 1.1171\neighborchange{\mathord{\uparrow}30.52}
 & 0.8569\neighborchange{\mathord{\uparrow}23.67}
 & \cellcolor{rankthird}\textit{0.087710}\neighborchange{\mathord{\downarrow}42.72}
 & 0.9846\neighborchange{\mathord{\uparrow}4.70}
 & 0.8664\neighborchange{\mathord{\uparrow}3.48}
 & 0.8156\neighborchange{\mathord{\uparrow}4.62}
 & 0.8103\neighborchange{\mathord{\uparrow}3.12} \\
PROD
 & 0.230181\neighborchange{\mathord{\downarrow}52.51}
 & \cellcolor{rankthird}\textit{0.7888}\neighborchange{\mathord{\uparrow}6.16}
 & \cellcolor{rankthird}\textit{0.6971}\neighborchange{\mathord{\uparrow}3.61}
 & \cellcolor{rankthird}\textit{0.9158}\neighborchange{\mathord{\uparrow}7.00}
 & \cellcolor{rankthird}\textit{0.7299}\neighborchange{\mathord{\uparrow}5.34}
 & 0.137420\neighborchange{\mathord{\downarrow}10.26}
 & \cellcolor{ranksecond}\underline{0.9503}\neighborchange{\mathord{\uparrow}1.05}
 & \cellcolor{rankfirst}\textbf{0.8403}\neighborchange{\mathord{\uparrow}0.36}
 & \cellcolor{ranksecond}\underline{0.7854}\neighborchange{\mathord{\uparrow}0.74}
 & \cellcolor{ranksecond}\underline{0.7886}\neighborchange{\mathord{\uparrow}0.36} \\
SimNPO-KL
 & 0.09843\neighborchange{\mathord{\downarrow}79.69}
 & 0.81680\neighborchange{\mathord{\uparrow}9.93}
 & 0.71245\neighborchange{\mathord{\uparrow}5.89}
 & 0.94280\neighborchange{\mathord{\uparrow}10.15}
 & 0.74802\neighborchange{\mathord{\uparrow}7.95}
 & 0.09791\neighborchange{\mathord{\downarrow}36.06}
 & 0.95822\neighborchange{\mathord{\uparrow}1.89}
 & 0.84386\neighborchange{\mathord{\uparrow}0.78}
 & 0.79081\neighborchange{\mathord{\uparrow}1.44}
 & 0.79263\neighborchange{\mathord{\uparrow}0.87} \\
Task Vector
 & 0.249535\neighborchange{\mathord{\downarrow}48.52}
 & \cellcolor{ranksecond}\underline{0.7790}\neighborchange{\mathord{\uparrow}4.85}
 & \cellcolor{ranksecond}\underline{0.6875}\neighborchange{\mathord{\uparrow}2.18}
 & \cellcolor{ranksecond}\underline{0.8919}\neighborchange{\mathord{\uparrow}4.21}
 & \cellcolor{ranksecond}\underline{0.7125}\neighborchange{\mathord{\uparrow}2.83}
 & 0.128422\neighborchange{\mathord{\downarrow}16.14}
 & \cellcolor{rankthird}\textit{0.9532}\neighborchange{\mathord{\uparrow}1.36}
 & \cellcolor{rankthird}\textit{0.8425}\neighborchange{\mathord{\uparrow}0.62}
 & \cellcolor{rankthird}\textit{0.7860}\neighborchange{\mathord{\uparrow}0.82}
 & \cellcolor{rankthird}\textit{0.7913}\neighborchange{\mathord{\uparrow}0.70} \\
GSS
 & \cellcolor{rankthird}\textit{0.09204}\neighborchange{\mathord{\downarrow}81.01}
 & 0.84599\neighborchange{\mathord{\uparrow}13.86}
 & 0.75151\neighborchange{\mathord{\uparrow}11.70}
 & 0.94035\neighborchange{\mathord{\uparrow}9.87}
 & 0.77645\neighborchange{\mathord{\uparrow}12.06}
 & \cellcolor{ranksecond}\underline{0.00620}\neighborchange{\mathord{\downarrow}95.95}
 & 3.12960\neighborchange{\mathord{\uparrow}232.79}
 & 2.13220\neighborchange{\mathord{\uparrow}154.65}
 & 2.02275\neighborchange{\mathord{\uparrow}159.46}
 & 1.76969\neighborchange{\mathord{\uparrow}125.21} \\
\midrule
\textbf{UNBIND}
 & \cellcolor{rankfirst}\textbf{0.004225}\neighborchange{\mathord{\downarrow}99.13}
 & \cellcolor{rankfirst}\textbf{0.7442}\neighborchange{\mathord{\uparrow}0.16}
 & \cellcolor{rankfirst}\textbf{0.6728}\neighborchange{\mathord{\approx}0}
 & \cellcolor{rankfirst}\textbf{0.8559}\neighborchange{\mathord{\approx}0}
 & \cellcolor{rankfirst}\textbf{0.6930}\neighborchange{\mathord{\uparrow}0.01}
 & \cellcolor{rankfirst}\textbf{0.004182}\neighborchange{\mathord{\downarrow}97.27}
 & \cellcolor{rankfirst}\textbf{0.9404}\neighborchange{\mathord{\approx}0}
 & \cellcolor{ranksecond}\underline{0.8415}\neighborchange{\mathord{\uparrow}0.50}
 & \cellcolor{rankfirst}\textbf{0.7796}\neighborchange{\mathord{\approx}0}
 & \cellcolor{rankfirst}\textbf{0.7880}\neighborchange{\mathord{\uparrow}0.28} \\
\bottomrule
\end{tabular}%
}
\end{table}
\vspace{-0.75em}

Table~\ref{tab:neighbors} shows that UNBIND reduces F-BLEU by 99.13\% on QwenCSN and 97.27\% on QwenStack, while limiting NLL increases across the four retain groups to 0.16\% and 0.50\%. These results show strong suppression of target code reproduction with little impact on related code. On CLStack, same-language and cross-language NLL increase by 2.3\% and 2.5\%, with respective 95\% intervals of 0.63\%--4.3\% and 0.32\%--5.9\%.
\FloatBarrier
\vspace{-0.75em}
\section{Conclusion}
\vspace{-0.5em}
We introduce UNBIND, which reframes code unlearning as selective representation intervention rather than parameter updating. Its key idea is to separate identifying target-related hidden states from determining how those states should be changed, and to apply the resulting intervention only when needed. This design enables strong forgetting while keeping model weights fixed and largely preserving general code capability. Across two models and two corpora, UNBIND reduces target reproduction by 97--99\%, achieves the strongest joint forgetting--utility performance among 14 baselines, and nearly eliminates long exact-span recovery under repeated extraction attempts. Multilingual and related-code evaluations further show that this suppression remains targeted rather than broadly degrading useful code representations.

\clearpage
\subsection*{AI use statement}
Generative AI tools assisted literature discovery, organization of the method and experimental evidence, manuscript writing, translation, and preparation of scripts for tables and figures. Numerical observations were taken from existing experiment records; manuscript preparation did not generate new model-evaluation results. The mathematical construction was checked against the method description, and displayed tables were checked against the reported observations. AI assistance also informed the presentation of design rationale and result interpretation. The authors are responsible for the final scientific claims, references, and artifacts.

\subsection*{Ethics statement}
This work studies suppression of designated code continuations using code corpora and programming benchmarks. Dataset use and any subsequent release of code examples must respect the applicable licenses and access restrictions.

\subsection*{Reproducibility statement}
The main text defines the intervention and evaluation protocol. Appendices~\ref{app:protocol}, \ref{app:construction},
and~\ref{app:baselines} provide data and adaptation settings,
UNBIND construction and calibration, and baseline implementations,
respectively. Appendices~\ref{app:access} and~\ref{app:multilingual}
detail the target-reproduction and multilingual evaluations. Our code and datasets are available at \url{https://anonymous.4open.science/r/UNBIND-75D0}.

\bibliographystyle{iclr2027_conference}
\bibliography{ref_acc}

@article{maini2024tofu,
  author       = {Pratyush Maini and
                  Zhili Feng and
                  Avi Schwarzschild and
                  Zachary C. Lipton and
                  J. Zico Kolter},
  title        = {{TOFU:} {A} Task of Fictitious Unlearning for LLMs},
  journal      = {CoRR},
  volume       = {abs/2401.06121},
  year         = {2024},
  url          = {https://doi.org/10.48550/arXiv.2401.06121},
  doi          = {10.48550/ARXIV.2401.06121},
  eprinttype   = {arXiv},
  eprint       = {2401.06121},
  bibsource    = {dblp computer science bibliography, https://dblp.org}
}

@article{zhang2024npo,
  author       = {Ruiqi Zhang and
                  Licong Lin and
                  Yu Bai and
                  Song Mei},
  title        = {Negative Preference Optimization: From Catastrophic Collapse to Effective
                  Unlearning},
  journal      = {CoRR},
  volume       = {abs/2404.05868},
  year         = {2024},
  url          = {https://doi.org/10.48550/arXiv.2404.05868},
  doi          = {10.48550/ARXIV.2404.05868},
  eprinttype   = {arXiv},
  eprint       = {2404.05868},
  bibsource    = {dblp computer science bibliography, https://dblp.org}
}

@inproceedings{fan2025simnpo,
  author       = {Chongyu Fan and
                  Jiancheng Liu and
                  Licong Lin and
                  Jinghan Jia and
                  Ruiqi Zhang and
                  Song Mei and
                  Sijia Liu},
  editor       = {Danielle Belgrave and
                  Cheng Zhang and
                  Laura N. Montoya and
                  Hsuan{-}Tien Lin and
                  Razvan Pascanu and
                  Piotr Koniusz and
                  Marzyeh Ghassemi and
                  Nancy Chen and
                  Iv{\'{a}}n Vladimir Meza Ru{\'{\i}}z and
                  Arturo Loaiza{-}Bonilla},
  title        = {Simplicity Prevails: Rethinking Negative Preference Optimization for
                  {LLM} Unlearning},
  booktitle    = {Advances in Neural Information Processing Systems 38: Annual Conference
                  on Neural Information Processing Systems 2025, NeurIPS 2025, San Diego,
                  CA, USA, December 2-7, 2025 / Mexico City, Mexico, November 30 - December
                  5, 2025},
  year         = {2025},
  url          = {http://papers.nips.cc/paper\_files/paper/2025/hash/02443e4e008231e0af0855f3cc70ed17-Abstract-Conference.html},
  bibsource    = {dblp computer science bibliography, https://dblp.org}
}

@inproceedings{rafailov2024dpov3,
  author       = {Rafael Rafailov and
                  Archit Sharma and
                  Eric Mitchell and
                  Christopher D. Manning and
                  Stefano Ermon and
                  Chelsea Finn},
  editor       = {Alice Oh and
                  Tristan Naumann and
                  Amir Globerson and
                  Kate Saenko and
                  Moritz Hardt and
                  Sergey Levine},
  title        = {Direct Preference Optimization: Your Language Model is Secretly a
                  Reward Model},
  booktitle    = {Advances in Neural Information Processing Systems 36: Annual Conference
                  on Neural Information Processing Systems 2023, NeurIPS 2023, New Orleans,
                  LA, USA, December 10 - 16, 2023},
  year         = {2023},
  url          = {http://papers.nips.cc/paper\_files/paper/2023/hash/a85b405ed65c6477a4fe8302b5e06ce7-Abstract-Conference.html},
  bibsource    = {dblp computer science bibliography, https://dblp.org}
}

@inproceedings{wang2025flat,
  author       = {Yaxuan Wang and
                  Jiaheng Wei and
                  Chris Yuhao Liu and
                  Jinlong Pang and
                  Quan Liu and
                  Ankit Shah and
                  Yujia Bao and
                  Yang Liu and
                  Wei Wei},
  title        = {{LLM} Unlearning via Loss Adjustment with Only Forget Data},
  booktitle    = {The Thirteenth International Conference on Learning Representations,
                  {ICLR} 2025, Singapore, April 24-28, 2025},
  publisher    = {OpenReview.net},
  year         = {2025},
  url          = {https://openreview.net/forum?id=6ESRicalFE},
  bibsource    = {dblp computer science bibliography, https://dblp.org}
}

@inproceedings{jiang2026prod,
  author       = {Xue Jiang and
                  Yihong Dong and
                  Huangzhao Zhang and
                  Tangxinyu Wang and
                  Zheng Fang and
                  Yingwei Ma and
                  Rongyu Cao and
                  Binhua Li and
                  Zhi Jin and
                  Wenpin Jiao and
                  Yongbin Li and
                  Ge Li},
  editor       = {Sven Koenig and
                  Chad Jenkins and
                  Matthew E. Taylor},
  title        = {Large Language Model Unlearning for Source Code},
  booktitle    = {Fortieth {AAAI} Conference on Artificial Intelligence, Thirty-Eighth
                  Conference on Innovative Applications of Artificial Intelligence,
                  Sixteenth Symposium on Educational Advances in Artificial Intelligence,
                  {AAAI} 2026, Singapore, January 20-27, 2026},
  pages        = {31346--31355},
  publisher    = {{AAAI} Press},
  year         = {2026},
  url          = {https://doi.org/10.1609/aaai.v40i37.40398},
  doi          = {10.1609/AAAI.V40I37.40398},
  bibsource    = {dblp computer science bibliography, https://dblp.org}
}

@inproceedings{li2024wmdpv7,
  author       = {Nathaniel Li and
                  Alexander Pan and
                  Anjali Gopal and
                  Summer Yue and
                  Daniel Berrios and
                  Alice Gatti and
                  Justin D. Li and
                  Ann{-}Kathrin Dombrowski and
                  Shashwat Goel and
                  Gabriel Mukobi and
                  Nathan Helm{-}Burger and
                  Rassin Lababidi and
                  Lennart Justen and
                  Andrew B. Liu and
                  Michael Chen and
                  Isabelle Barrass and
                  Oliver Zhang and
                  Xiaoyuan Zhu and
                  Rishub Tamirisa and
                  Bhrugu Bharathi and
                  Ariel Herbert{-}Voss and
                  Cort B. Breuer and
                  Andy Zou and
                  Mantas Mazeika and
                  Zifan Wang and
                  Palash Oswal and
                  Weiran Lin and
                  Adam A. Hunt and
                  Justin Tienken{-}Harder and
                  Kevin Y. Shih and
                  Kemper Talley and
                  John Guan and
                  Ian Steneker and
                  David Campbell and
                  Brad Jokubaitis and
                  Steven Basart and
                  Stephen Fitz and
                  Ponnurangam Kumaraguru and
                  Kallol Krishna Karmakar and
                  Uday Kiran Tupakula and
                  Vijay Varadharajan and
                  Yan Shoshitaishvili and
                  Jimmy Ba and
                  Kevin M. Esvelt and
                  Alexandr Wang and
                  Dan Hendrycks},
  editor       = {Ruslan Salakhutdinov and
                  Zico Kolter and
                  Katherine A. Heller and
                  Adrian Weller and
                  Nuria Oliver and
                  Jonathan Scarlett and
                  Felix Berkenkamp},
  title        = {The {WMDP} Benchmark: Measuring and Reducing Malicious Use with Unlearning},
  booktitle    = {Forty-first International Conference on Machine Learning, {ICML} 2024,
                  Vienna, Austria, July 21-27, 2024},
  series       = {Proceedings of Machine Learning Research},
  volume       = {235},
  pages        = {28525--28550},
  publisher    = {{PMLR} / OpenReview.net},
  year         = {2024},
  url          = {https://proceedings.mlr.press/v235/li24bc.html},
  bibsource    = {dblp computer science bibliography, https://dblp.org}
}

@article{zarezade2026asu,
  author       = {Saleh Zare Zade and
                  Xiangyu Zhou and
                  Sijia Liu and
                  Dongxiao Zhu},
  title        = {Attention Smoothing Is All You Need For Unlearning},
  journal      = {CoRR},
  volume       = {abs/2603.01285},
  year         = {2026},
  url          = {https://doi.org/10.48550/arXiv.2603.01285},
  doi          = {10.48550/ARXIV.2603.01285},
  eprinttype   = {arXiv},
  eprint       = {2603.01285},
  bibsource    = {dblp computer science bibliography, https://dblp.org}
}

@article{hui2024qwen25coder,
  author       = {Binyuan Hui and
                  Jian Yang and
                  Zeyu Cui and
                  Jiaxi Yang and
                  Dayiheng Liu and
                  Lei Zhang and
                  Tianyu Liu and
                  Jiajun Zhang and
                  Bowen Yu and
                  Kai Dang and
                  An Yang and
                  Rui Men and
                  Fei Huang and
                  Xingzhang Ren and
                  Xuancheng Ren and
                  Jingren Zhou and
                  Junyang Lin},
  title        = {Qwen2.5-Coder Technical Report},
  journal      = {CoRR},
  volume       = {abs/2409.12186},
  year         = {2024},
  url          = {https://doi.org/10.48550/arXiv.2409.12186},
  doi          = {10.48550/ARXIV.2409.12186},
  eprinttype   = {arXiv},
  eprint       = {2409.12186},
  bibsource    = {dblp computer science bibliography, https://dblp.org}
}

@article{roziere2023codellama,
  author       = {Baptiste Rozi{\`{e}}re and
                  Jonas Gehring and
                  Fabian Gloeckle and
                  Sten Sootla and
                  Itai Gat and
                  Xiaoqing Ellen Tan and
                  Yossi Adi and
                  Jingyu Liu and
                  Tal Remez and
                  J{\'{e}}r{\'{e}}my Rapin and
                  Artyom Kozhevnikov and
                  Ivan Evtimov and
                  Joanna Bitton and
                  Manish Bhatt and
                  Cristian Canton{-}Ferrer and
                  Aaron Grattafiori and
                  Wenhan Xiong and
                  Alexandre D{\'{e}}fossez and
                  Jade Copet and
                  Faisal Azhar and
                  Hugo Touvron and
                  Louis Martin and
                  Nicolas Usunier and
                  Thomas Scialom and
                  Gabriel Synnaeve},
  title        = {Code Llama: Open Foundation Models for Code},
  journal      = {CoRR},
  volume       = {abs/2308.12950},
  year         = {2023},
  url          = {https://doi.org/10.48550/arXiv.2308.12950},
  doi          = {10.48550/ARXIV.2308.12950},
  eprinttype   = {arXiv},
  eprint       = {2308.12950},
  bibsource    = {dblp computer science bibliography, https://dblp.org}
}

@inproceedings{liu2023evalplus,
  author       = {Jiawei Liu and
                  Chunqiu Steven Xia and
                  Yuyao Wang and
                  Lingming Zhang},
  editor       = {Alice Oh and
                  Tristan Naumann and
                  Amir Globerson and
                  Kate Saenko and
                  Moritz Hardt and
                  Sergey Levine},
  title        = {Is Your Code Generated by ChatGPT Really Correct? Rigorous Evaluation
                  of Large Language Models for Code Generation},
  booktitle    = {Advances in Neural Information Processing Systems 36: Annual Conference
                  on Neural Information Processing Systems 2023, NeurIPS 2023, New Orleans,
                  LA, USA, December 10 - 16, 2023},
  year         = {2023},
  url          = {http://papers.nips.cc/paper\_files/paper/2023/hash/43e9d647ccd3e4b7b5baab53f0368686-Abstract-Conference.html},
  bibsource    = {dblp computer science bibliography, https://dblp.org}
}

@article{li2023starcoder,
  author       = {Raymond Li and
                  Loubna Ben Allal and
                  Yangtian Zi and
                  Niklas Muennighoff and
                  Denis Kocetkov and
                  Chenghao Mou and
                  Marc Marone and
                  Christopher Akiki and
                  Jia Li and
                  Jenny Chim and
                  Qian Liu and
                  Evgenii Zheltonozhskii and
                  Terry Yue Zhuo and
                  Thomas Wang and
                  Olivier Dehaene and
                  Mishig Davaadorj and
                  Joel Lamy{-}Poirier and
                  Jo{\~{a}}o Monteiro and
                  Oleh Shliazhko and
                  Nicolas Gontier and
                  Nicholas Meade and
                  Armel Zebaze and
                  Ming{-}Ho Yee and
                  Logesh Kumar Umapathi and
                  Jian Zhu and
                  Benjamin Lipkin and
                  Muhtasham Oblokulov and
                  Zhiruo Wang and
                  Rudra Murthy V and
                  Jason T. Stillerman and
                  Siva Sankalp Patel and
                  Dmitry Abulkhanov and
                  Marco Zocca and
                  Manan Dey and
                  Zhihan Zhang and
                  Nour Fahmy and
                  Urvashi Bhattacharyya and
                  Wenhao Yu and
                  Swayam Singh and
                  Sasha Luccioni and
                  Paulo Villegas and
                  Maxim Kunakov and
                  Fedor Zhdanov and
                  Manuel Romero and
                  Tony Lee and
                  Nadav Timor and
                  Jennifer Ding and
                  Claire Schlesinger and
                  Hailey Schoelkopf and
                  Jan Ebert and
                  Tri Dao and
                  Mayank Mishra and
                  Alex Gu and
                  Jennifer Robinson and
                  Carolyn Jane Anderson and
                  Brendan Dolan{-}Gavitt and
                  Danish Contractor and
                  Siva Reddy and
                  Daniel Fried and
                  Dzmitry Bahdanau and
                  Yacine Jernite and
                  Carlos Mu{\~{n}}oz Ferrandis and
                  Sean Hughes and
                  Thomas Wolf and
                  Arjun Guha and
                  Leandro von Werra and
                  Harm de Vries},
  title        = {StarCoder: may the source be with you!},
  journal      = {Trans. Mach. Learn. Res.},
  volume       = {2023},
  year         = {2023},
  url          = {https://openreview.net/forum?id=KoFOg41haE},
  bibsource    = {dblp computer science bibliography, https://dblp.org}
}

@article{zou2023representationengineering,
  author       = {Andy Zou and
                  Long Phan and
                  Sarah Li Chen and
                  James Campbell and
                  Phillip Guo and
                  Richard Ren and
                  Alexander Pan and
                  Xuwang Yin and
                  Mantas Mazeika and
                  Ann{-}Kathrin Dombrowski and
                  Shashwat Goel and
                  Nathaniel Li and
                  Michael J. Byun and
                  Zifan Wang and
                  Alex Mallen and
                  Steven Basart and
                  Sanmi Koyejo and
                  Dawn Song and
                  Matt Fredrikson and
                  J. Zico Kolter and
                  Dan Hendrycks},
  title        = {Representation Engineering: {A} Top-Down Approach to {AI} Transparency},
  journal      = {CoRR},
  volume       = {abs/2310.01405},
  year         = {2023},
  url          = {https://doi.org/10.48550/arXiv.2310.01405},
  doi          = {10.48550/ARXIV.2310.01405},
  eprinttype   = {arXiv},
  eprint       = {2310.01405},
  bibsource    = {dblp computer science bibliography, https://dblp.org}
}

@article{turner2023activationaddition,
  author       = {Alexander Matt Turner and
                  Lisa Thiergart and
                  David Udell and
                  Gavin Leech and
                  Ulisse Mini and
                  Monte MacDiarmid},
  title        = {Activation Addition: Steering Language Models Without Optimization},
  journal      = {CoRR},
  volume       = {abs/2308.10248},
  year         = {2023},
  url          = {https://doi.org/10.48550/arXiv.2308.10248},
  doi          = {10.48550/ARXIV.2308.10248},
  eprinttype   = {arXiv},
  eprint       = {2308.10248},
  bibsource    = {dblp computer science bibliography, https://dblp.org}
}

@inproceedings{li2023iti,
  author       = {Kenneth Li and
                  Oam Patel and
                  Fernanda B. Vi{\'{e}}gas and
                  Hanspeter Pfister and
                  Martin Wattenberg},
  editor       = {Alice Oh and
                  Tristan Naumann and
                  Amir Globerson and
                  Kate Saenko and
                  Moritz Hardt and
                  Sergey Levine},
  title        = {Inference-Time Intervention: Eliciting Truthful Answers from a Language
                  Model},
  booktitle    = {Advances in Neural Information Processing Systems 36: Annual Conference
                  on Neural Information Processing Systems 2023, NeurIPS 2023, New Orleans,
                  LA, USA, December 10 - 16, 2023},
  year         = {2023},
  url          = {http://papers.nips.cc/paper\_files/paper/2023/hash/81b8390039b7302c909cb769f8b6cd93-Abstract-Conference.html},
  bibsource    = {dblp computer science bibliography, https://dblp.org}
}

@inproceedings{arditi2024refusaldirection,
  author       = {Andy Arditi and
                  Oscar Obeso and
                  Aaquib Syed and
                  Daniel Paleka and
                  Nina Panickssery and
                  Wes Gurnee and
                  Neel Nanda},
  editor       = {Amir Globersons and
                  Lester Mackey and
                  Danielle Belgrave and
                  Angela Fan and
                  Ulrich Paquet and
                  Jakub M. Tomczak and
                  Cheng Zhang},
  title        = {Refusal in Language Models Is Mediated by a Single Direction},
  booktitle    = {Advances in Neural Information Processing Systems 37: Annual Conference
                  on Neural Information Processing Systems 2024, NeurIPS 2024, Vancouver,
                  BC, Canada, December 10 - 15, 2024},
  year         = {2024},
  url          = {http://papers.nips.cc/paper\_files/paper/2024/hash/f545448535dfde4f9786555403ab7c49-Abstract-Conference.html},
  bibsource    = {dblp computer science bibliography, https://dblp.org}
}

@article{husain2019codesearchnet,
  author       = {Hamel Husain and
                  Ho{-}Hsiang Wu and
                  Tiferet Gazit and
                  Miltiadis Allamanis and
                  Marc Brockschmidt},
  title        = {CodeSearchNet Challenge: Evaluating the State of Semantic Code Search},
  journal      = {CoRR},
  volume       = {abs/1909.09436},
  year         = {2019},
  url          = {http://arxiv.org/abs/1909.09436},
  eprinttype   = {arXiv},
  eprint       = {1909.09436},
  bibsource    = {dblp computer science bibliography, https://dblp.org}
}

@article{kocetkov2022stack,
  author       = {Denis Kocetkov and
                  Raymond Li and
                  Loubna Ben Allal and
                  Jia Li and
                  Chenghao Mou and
                  Yacine Jernite and
                  Margaret Mitchell and
                  Carlos Mu{\~{n}}oz Ferrandis and
                  Sean Hughes and
                  Thomas Wolf and
                  Dzmitry Bahdanau and
                  Leandro von Werra and
                  Harm de Vries},
  title        = {The Stack: 3 {TB} of permissively licensed source code},
  journal      = {Trans. Mach. Learn. Res.},
  volume       = {2023},
  year         = {2023},
  url          = {https://openreview.net/forum?id=pxpbTdUEpD},
  bibsource    = {dblp computer science bibliography, https://dblp.org}
}

@inproceedings{hu2022lora,
  author       = {Edward J. Hu and
                  Yelong Shen and
                  Phillip Wallis and
                  Zeyuan Allen{-}Zhu and
                  Yuanzhi Li and
                  Shean Wang and
                  Lu Wang and
                  Weizhu Chen},
  title        = {LoRA: Low-Rank Adaptation of Large Language Models},
  booktitle    = {The Tenth International Conference on Learning Representations, {ICLR}
                  2022, Virtual Event, April 25-29, 2022},
  publisher    = {OpenReview.net},
  year         = {2022},
  url          = {https://openreview.net/forum?id=nZeVKeeFYf9},
  bibsource    = {dblp computer science bibliography, https://dblp.org}
}

@article{chen2021humaneval,
  author       = {Mark Chen and
                  Jerry Tworek and
                  Heewoo Jun and
                  Qiming Yuan and
                  Henrique Pond{\'{e}} de Oliveira Pinto and
                  Jared Kaplan and
                  Harri Edwards and
                  Yuri Burda and
                  Nicholas Joseph and
                  Greg Brockman and
                  Alex Ray and
                  Raul Puri and
                  Gretchen Krueger and
                  Michael Petrov and
                  Heidy Khlaaf and
                  Girish Sastry and
                  Pamela Mishkin and
                  Brooke Chan and
                  Scott Gray and
                  Nick Ryder and
                  Mikhail Pavlov and
                  Alethea Power and
                  Lukasz Kaiser and
                  Mohammad Bavarian and
                  Clemens Winter and
                  Philippe Tillet and
                  Felipe Petroski Such and
                  Dave Cummings and
                  Matthias Plappert and
                  Fotios Chantzis and
                  Elizabeth Barnes and
                  Ariel Herbert{-}Voss and
                  William Hebgen Guss and
                  Alex Nichol and
                  Alex Paino and
                  Nikolas Tezak and
                  Jie Tang and
                  Igor Babuschkin and
                  Suchir Balaji and
                  Shantanu Jain and
                  William Saunders and
                  Christopher Hesse and
                  Andrew N. Carr and
                  Jan Leike and
                  Joshua Achiam and
                  Vedant Misra and
                  Evan Morikawa and
                  Alec Radford and
                  Matthew Knight and
                  Miles Brundage and
                  Mira Murati and
                  Katie Mayer and
                  Peter Welinder and
                  Bob McGrew and
                  Dario Amodei and
                  Sam McCandlish and
                  Ilya Sutskever and
                  Wojciech Zaremba},
  title        = {Evaluating Large Language Models Trained on Code},
  journal      = {CoRR},
  volume       = {abs/2107.03374},
  year         = {2021},
  url          = {https://arxiv.org/abs/2107.03374},
  eprinttype   = {arXiv},
  eprint       = {2107.03374},
  bibsource    = {dblp computer science bibliography, https://dblp.org}
}

@article{austin2021mbpp,
  author       = {Jacob Austin and
                  Augustus Odena and
                  Maxwell I. Nye and
                  Maarten Bosma and
                  Henryk Michalewski and
                  David Dohan and
                  Ellen Jiang and
                  Carrie J. Cai and
                  Michael Terry and
                  Quoc V. Le and
                  Charles Sutton},
  title        = {Program Synthesis with Large Language Models},
  journal      = {CoRR},
  volume       = {abs/2108.07732},
  year         = {2021},
  url          = {https://arxiv.org/abs/2108.07732},
  eprinttype   = {arXiv},
  eprint       = {2108.07732},
  bibsource    = {dblp computer science bibliography, https://dblp.org}
}

@inproceedings{guo2022unixcoder,
  author       = {Daya Guo and
                  Shuai Lu and
                  Nan Duan and
                  Yanlin Wang and
                  Ming Zhou and
                  Jian Yin},
  editor       = {Smaranda Muresan and
                  Preslav Nakov and
                  Aline Villavicencio},
  title        = {UniXcoder: Unified Cross-Modal Pre-training for Code Representation},
  booktitle    = {Proceedings of the 60th Annual Meeting of the Association for Computational
                  Linguistics (Volume 1: Long Papers), {ACL} 2022, Dublin, Ireland,
                  May 22-27, 2022},
  pages        = {7212--7225},
  publisher    = {Association for Computational Linguistics},
  year         = {2022},
  url          = {https://doi.org/10.18653/v1/2022.acl-long.499},
  doi          = {10.18653/V1/2022.ACL-LONG.499},
  bibsource    = {dblp computer science bibliography, https://dblp.org}
}

@inproceedings{papineni2002bleu,
  author       = {Kishore Papineni and
                  Salim Roukos and
                  Todd Ward and
                  Wei{-}Jing Zhu},
  title        = {Bleu: a Method for Automatic Evaluation of Machine Translation},
  booktitle    = {Proceedings of the 40th Annual Meeting of the Association for Computational
                  Linguistics, July 6-12, 2002, Philadelphia, PA, {USA}},
  pages        = {311--318},
  publisher    = {{ACL}},
  year         = {2002},
  url          = {https://aclanthology.org/P02-1040/},
  doi          = {10.3115/1073083.1073135},
  bibsource    = {dblp computer science bibliography, https://dblp.org}
}

@inproceedings{cao2015unlearning,
  author       = {Yinzhi Cao and
                  Junfeng Yang},
  title        = {Towards Making Systems Forget with Machine Unlearning},
  booktitle    = {2015 {IEEE} Symposium on Security and Privacy, {SP} 2015, San Jose,
                  CA, USA, May 17-21, 2015},
  pages        = {463--480},
  publisher    = {{IEEE} Computer Society},
  year         = {2015},
  url          = {https://doi.org/10.1109/SP.2015.35},
  doi          = {10.1109/SP.2015.35},
  bibsource    = {dblp computer science bibliography, https://dblp.org}
}

@inproceedings{bourtoule2021unlearning,
  author       = {Lucas Bourtoule and
                  Varun Chandrasekaran and
                  Christopher A. Choquette{-}Choo and
                  Hengrui Jia and
                  Adelin Travers and
                  Baiwu Zhang and
                  David Lie and
                  Nicolas Papernot},
  title        = {Machine Unlearning},
  booktitle    = {42nd {IEEE} Symposium on Security and Privacy, {SP} 2021, San Francisco,
                  CA, USA, 24-27 May 2021},
  pages        = {141--159},
  publisher    = {{IEEE}},
  year         = {2021},
  url          = {https://doi.org/10.1109/SP40001.2021.00019},
  doi          = {10.1109/SP40001.2021.00019},
  bibsource    = {dblp computer science bibliography, https://dblp.org}
}

@inproceedings{guo2020certified,
  author       = {Chuan Guo and
                  Tom Goldstein and
                  Awni Y. Hannun and
                  Laurens van der Maaten},
  title        = {Certified Data Removal from Machine Learning Models},
  booktitle    = {Proceedings of the 37th International Conference on Machine Learning,
                  {ICML} 2020, 13-18 July 2020, Virtual Event},
  series       = {Proceedings of Machine Learning Research},
  volume       = {119},
  pages        = {3832--3842},
  publisher    = {{PMLR}},
  year         = {2020},
  url          = {http://proceedings.mlr.press/v119/guo20c.html},
  bibsource    = {dblp computer science bibliography, https://dblp.org}
}

@inproceedings{carlini2021extracting,
  author       = {Nicholas Carlini and
                  Florian Tram{\`{e}}r and
                  Eric Wallace and
                  Matthew Jagielski and
                  Ariel Herbert{-}Voss and
                  Katherine Lee and
                  Adam Roberts and
                  Tom B. Brown and
                  Dawn Song and
                  {\'{U}}lfar Erlingsson and
                  Alina Oprea and
                  Colin Raffel},
  editor       = {Michael D. Bailey and
                  Rachel Greenstadt},
  title        = {Extracting Training Data from Large Language Models},
  booktitle    = {30th {USENIX} Security Symposium, {USENIX} Security 2021, August 11-13,
                  2021},
  pages        = {2633--2650},
  publisher    = {{USENIX} Association},
  year         = {2021},
  url          = {https://www.usenix.org/conference/usenixsecurity21/presentation/carlini-extracting},
  bibsource    = {dblp computer science bibliography, https://dblp.org}
}

@inproceedings{carlini2023quantifying,
  author       = {Nicholas Carlini and
                  Daphne Ippolito and
                  Matthew Jagielski and
                  Katherine Lee and
                  Florian Tram{\`{e}}r and
                  Chiyuan Zhang},
  title        = {Quantifying Memorization Across Neural Language Models},
  booktitle    = {The Eleventh International Conference on Learning Representations,
                  {ICLR} 2023, Kigali, Rwanda, May 1-5, 2023},
  publisher    = {OpenReview.net},
  year         = {2023},
  url          = {https://openreview.net/forum?id=TatRHT\_1cK},
  bibsource    = {dblp computer science bibliography, https://dblp.org}
}

@inproceedings{lee2022deduplicating,
  author       = {Katherine Lee and
                  Daphne Ippolito and
                  Andrew Nystrom and
                  Chiyuan Zhang and
                  Douglas Eck and
                  Chris Callison{-}Burch and
                  Nicholas Carlini},
  editor       = {Smaranda Muresan and
                  Preslav Nakov and
                  Aline Villavicencio},
  title        = {Deduplicating Training Data Makes Language Models Better},
  booktitle    = {Proceedings of the 60th Annual Meeting of the Association for Computational
                  Linguistics (Volume 1: Long Papers), {ACL} 2022, Dublin, Ireland,
                  May 22-27, 2022},
  pages        = {8424--8445},
  publisher    = {Association for Computational Linguistics},
  year         = {2022},
  url          = {https://doi.org/10.18653/v1/2022.acl-long.577},
  doi          = {10.18653/V1/2022.ACL-LONG.577},
  bibsource    = {dblp computer science bibliography, https://dblp.org}
}

@inproceedings{pawelczyk2024incontext,
  author       = {Martin Pawelczyk and
                  Seth Neel and
                  Himabindu Lakkaraju},
  editor       = {Ruslan Salakhutdinov and
                  Zico Kolter and
                  Katherine A. Heller and
                  Adrian Weller and
                  Nuria Oliver and
                  Jonathan Scarlett and
                  Felix Berkenkamp},
  title        = {In-Context Unlearning: Language Models as Few-Shot Unlearners},
  booktitle    = {Forty-first International Conference on Machine Learning, {ICML} 2024,
                  Vienna, Austria, July 21-27, 2024},
  series       = {Proceedings of Machine Learning Research},
  volume       = {235},
  pages        = {40034--40050},
  publisher    = {{PMLR} / OpenReview.net},
  year         = {2024},
  url          = {https://proceedings.mlr.press/v235/pawelczyk24a.html},
  bibsource    = {dblp computer science bibliography, https://dblp.org}
}

@inproceedings{rimsky2024caa,
  author       = {Nina Rimsky and
                  Nick Gabrieli and
                  Julian Schulz and
                  Meg Tong and
                  Evan Hubinger and
                  Alexander Matt Turner},
  editor       = {Lun{-}Wei Ku and
                  Andre Martins and
                  Vivek Srikumar},
  title        = {Steering Llama 2 via Contrastive Activation Addition},
  booktitle    = {Proceedings of the 62nd Annual Meeting of the Association for Computational
                  Linguistics (Volume 1: Long Papers), {ACL} 2024, Bangkok, Thailand,
                  August 11-16, 2024},
  pages        = {15504--15522},
  publisher    = {Association for Computational Linguistics},
  year         = {2024},
  url          = {https://doi.org/10.18653/v1/2024.acl-long.828},
  doi          = {10.18653/V1/2024.ACL-LONG.828},
  bibsource    = {dblp computer science bibliography, https://dblp.org}
}

@article{zhang2026gssgatedsubspacesteering,
  author       = {Xuanqi Zhang and
                  Haoyang Shang and
                  Xiaoxiao Li},
  title        = {{GSS:} Gated Subspace Steering for Selective Memorization Mitigation
                  in LLMs},
  journal      = {CoRR},
  volume       = {abs/2602.08901},
  year         = {2026},
  url          = {https://doi.org/10.48550/arXiv.2602.08901},
  doi          = {10.48550/ARXIV.2602.08901},
  eprinttype   = {arXiv},
  eprint       = {2602.08901},
  bibsource    = {dblp computer science bibliography, https://dblp.org}
}

@article{shan2026sage,
  author       = {Zhengyang Shan and
                  Xu Qian and
                  Jiayun Xin and
                  Minghui Xu and
                  Yue Zhang and
                  Zhen Yang and
                  Hao Wu and
                  Xiuzhen Cheng},
  title        = {{SAGE:} Signal-Amplified Guided Embeddings for LLM-based Vulnerability
                  Detection},
  journal      = {CoRR},
  volume       = {abs/2604.19031},
  year         = {2026},
  url          = {https://doi.org/10.48550/arXiv.2604.19031},
  doi          = {10.48550/ARXIV.2604.19031},
  eprinttype   = {arXiv},
  eprint       = {2604.19031},
  bibsource    = {dblp computer science bibliography, https://dblp.org}
}

@article{chu2026codeeraser,
  author       = {Zhaoyang Chu and
                  Yao Wan and
                  Zhikun Zhang and
                  Di Wang and
                  Zhou Yang and
                  Hongyu Zhang and
                  Pan Zhou and
                  Xuanhua Shi and
                  Hai Jin and
                  David Lo},
  title        = {Scrub It Out! Erasing Sensitive Memorization in Code Language Models
                  via Machine Unlearning},
  journal      = {CoRR},
  volume       = {abs/2509.13755},
  year         = {2025},
  url          = {https://doi.org/10.48550/arXiv.2509.13755},
  doi          = {10.48550/ARXIV.2509.13755},
  eprinttype   = {arXiv},
  eprint       = {2509.13755},
  bibsource    = {dblp computer science bibliography, https://dblp.org}
}

@article{chowdhury2026conformal,
  author       = {Somnath Basu Roy Chowdhury and
                  Rahul Kidambi and
                  Avinava Dubey and
                  David Wang and
                  G{\"{o}}khan Mergen and
                  Amr Ahmed and
                  Aranyak Mehta},
  title        = {Inference-time Unlearning Using Conformal Prediction},
  journal      = {CoRR},
  volume       = {abs/2602.03787},
  year         = {2026},
  url          = {https://doi.org/10.48550/arXiv.2602.03787},
  doi          = {10.48550/ARXIV.2602.03787},
  eprinttype   = {arXiv},
  eprint       = {2602.03787},
  bibsource    = {dblp computer science bibliography, https://dblp.org}
}

@article{merchant2026divergence,
  author       = {Humzah Merchant and
                  Bradford Levy},
  title        = {Divergence Decoding: Inference-Time Unlearning via Auxiliary Models},
  journal      = {CoRR},
  volume       = {abs/2605.31293},
  year         = {2026},
  url          = {https://doi.org/10.48550/arXiv.2605.31293},
  doi          = {10.48550/ARXIV.2605.31293},
  eprinttype   = {arXiv},
  eprint       = {2605.31293},
  bibsource    = {dblp computer science bibliography, https://dblp.org}
}

@article{chen2026st2u,
  author       = {Xunlei Chen and
                  Qinghui Gong and
                  Ruini Xue and
                  Yaodong Hu and
                  Tian Lan and
                  Wenhong Tian},
  title        = {ST\({}^{\mbox{2}}\)U: Stateful Test-Time Unlearning via Restricted
                  Knowledge Boundary Control},
  journal      = {CoRR},
  volume       = {abs/2608.23034},
  year         = {2026},
  url          = {https://doi.org/10.48550/arXiv.2608.23034},
  doi          = {10.48550/ARXIV.2608.23034},
  eprinttype   = {arXiv},
  eprint       = {2608.23034},
  bibsource    = {dblp computer science bibliography, https://dblp.org}
}
\clearpage
\appendix
\section{Experimental Protocol}
\label{app:protocol}

\subsection{Data Construction and Splits}
\label{app:protocol:data}

We use the frozen seed-0 CodeSearchNet (CSN) and The Stack v1
9,600-example manifests. Each corpus contains 300/50/50 forget
training/validation/test examples and 8,280/460/460 retain examples.
The main forget targets are Python code. Retain data cover Python,
Go, Java, JavaScript, PHP, and Ruby for CSN, and Python, C++, Go,
Java, JavaScript, and Rust for The Stack. Existing source records,
ordering, and split assignments are reused without resampling.

\begin{table*}[t]
\centering
\small
\renewcommand{\arraystretch}{1.1}
\caption{Data roles in the experiments. All partitions follow the frozen
seed-0 CSN and The Stack manifests and are assigned by source row.}
\label{tab:protocol:data}
\begin{tabularx}{\textwidth}{@{}l l X@{}}
\toprule
Purpose & Size & Usage and overlap \\
\midrule

Adaptation
& 300 forget; 8,280 retain
& Training pool for the paired reference models. \\

Detector/output-axis fitting
& 240 forget; 480 retain
& Fitting subsets used to construct the detector and output axis;
disjoint from all calibration rows. \\

Forget calibration-fit
& 50 forget
& Used to construct candidate quantile thresholds $\tau_q$ and orient
the corresponding output directions $v_q$. \\

Forget calibration-validation
& 10 forget
& Held-out forget rows used for constrained $(q,\kappa)$ selection;
disjoint from the 240 fitting and 50 calibration-fit rows. \\

Retain calibration
& 120 retain
& Language-proportional retain subset used to measure KL, NLL, and
BLEU constraints during $(q,\kappa)$ selection. \\

Main evaluation
& 300 forget; 460 retain
& Reproduction of the designated forget-training targets;
retain likelihoods use retain-test. \\

Access evaluation
& 300 targets
& Reuses the main forget targets under multiple access conditions. \\

Multilingual evaluation
& 300 forget per language; 460 shared retain
& Separate Python, Java, and JavaScript forget tasks;
shared retain-test data and Python HE+/MBPP+ utility tests. \\

Related-code evaluation
& 512 per evaluated setting
& Four groups of 128: same-language semantic neighbors,
cross-language semantic neighbors, ordinary retain, and
format-matched retain. \\

\bottomrule
\end{tabularx}
\end{table*}

Construction partitions are assigned by source row, keeping all windows
from the same row together. The 300 forget rows are first split into
240 fitting rows and 60 calibration rows with seed 0. The latter are
further divided into 50 calibration-fit rows and 10
calibration-validation rows. The calibration-fit rows are used to
construct the candidate quantile thresholds $\tau_q$ and orient the
corresponding output directions $v_q$, while the
calibration-validation rows are used for constrained $(q,\kappa)$
selection. A language-proportional sample of 600 retain-training rows
is partitioned within language into fitting and calibration subsets,
yielding 480/120 rows for CSN and Stack.
The retain calibration rows provide the retain-side measurements used
to enforce the KL, NLL, and BLEU constraints during $(q,\kappa)$
selection. 

\paragraph{Calibration scope.}
For the reported operating point, the calibration sweep selects
$q^\star=0.25$ and $\kappa^\star=20$. For each candidate quantile
$q$, the threshold $\tau_q$ is computed by linear interpolation over
positive forget-calibration activations, with repeated values retained.
The sign of the output direction is oriented on the same
calibration-training rows using gate-weighted target-loss gradient
projections. The resulting $(\tau_q,v_q)$ pairs are then combined with
candidate strengths $\kappa\in\mathcal K$ and evaluated on a separate
calibration-validation partition. Candidates must satisfy the prescribed
retain KL, retain-NLL, and retain-BLEU budgets, after which the feasible
pair with the highest forget validation NLL is selected. Main F-BLEU
evaluates all 300 designated forget-training targets, including
calibration examples; it therefore measures reproduction of the
designated training targets rather than held-out forgetting generalization.

\subsection{Model Adaptation and Reference Models}
\label{app:protocol:models}

We use Qwen2.5-Coder-7B-Instruct and CodeLlama-7b-hf with their own
tokenizers. $M_{\mathrm{base}}$ denotes the checkpoint before adaptation;
$M_{\mathrm{full}}$ and $M_{\mathrm{retrain}}$ start from the same base
weights and saved seed-0 LoRA initialization, with the latter excluding
forget slots. LoRA uses rank 32, alpha 64, dropout 0, and the
\texttt{q\_proj}, \texttt{k\_proj}, \texttt{v\_proj},
\texttt{o\_proj}, \texttt{gate\_proj}, \texttt{up\_proj}, and
\texttt{down\_proj} modules. Training uses AdamW with constant learning
rate $5\times10^{-5}$, $(\beta_1,\beta_2)=(0.9,0.999)$,
$\epsilon=10^{-8}$, weight decay 0.01, gradient clipping at 1.0,
maximum length 2,048, BF16 forward computation, and FP32 trainable
master weights.

Each scheduled epoch interleaves 300 forget slots with 2,700 retain
slots drawn from a seeded, repeatedly shuffled retain stream.
Microbatch size is one; gradients accumulate over 32 scheduled slots
(24 in the final batch), with loss normalized by the number of scored
tokens. Both models perform 94 updates per epoch.
$M_{\mathrm{retrain}}$ skips forget slots without replacing them or
changing update boundaries. Retain exposure, update counts, and the
constant learning rate are consequently matched, although its number
of live examples per update is smaller.

The reported paired checkpoints are epoch 8 for Qwen/CSN, epoch 9
for Qwen/Stack, and epoch 10 for both CodeLlama settings, corresponding
to 752, 846, and 940 updates. These saved checkpoints are reused in
the rerun. We scan all 28 decoder layers of Qwen2.5-Coder-7B and all 32 decoder layers of CodeLlama-7B. The selected zero-based layers are 23, 24, 29, and 28 for QwenCSN, QwenStack, CLCSN, and CLStack, respectively. Construction and threshold–strength calibration are performed separately at each candidate layer, as described in Appendix B.3.

\subsection{Evaluation Metrics and Generation Settings}
\label{app:protocol:metrics}

\paragraph{Main continuation metrics.}
For the main F-BLEU and R-BLEU evaluation, we use one greedy
completion per example (\texttt{num\_beams=1}), BF16, batch size 4,
and at most 128 new tokens. Forget prompts and target continuations
are read directly from the frozen target manifest, without applying
a chat template. For each retain-test program, we choose the latest
tokenizer-offset boundary that leaves at least 32 prefix tokens; only the prefix is left-truncated when needed
to satisfy the 2,048-token composed-input limit. 

F-BLEU and R-BLEU are arithmetic means of sentence-level BLEU-4,
using the evaluated model's tokenizer, uniform n-gram weights, and
the standard brevity penalty. The preserved implementation replaces
a zero modified precision by $1/\max(2N_n,1)$, where $N_n$ is the
number of hypothesis $n$-grams; an empty reference or hypothesis
scores zero. Generated continuations are compared with the stored
reference continuations, excluding prompt text. Scores use the
$[0,1]$ scale and are not corpus-level BLEU.

R-PPL uses all 460 retain-test programs. Each sequence consists of
one BOS token (EOS when BOS is unavailable) followed by at most
2,047 native-code tokens. Only native-code target tokens contribute
to the loss; the prepended special token and padding positions are
masked. We exponentiate the total token NLL divided by the total
number of scored tokens, rather than averaging per-example
perplexities. CU uses a method-specific conditioning context for its reported retain likelihood; its prompt-conditioned target-suffix evaluation is described in Appendix C.3.

\paragraph{Functional correctness.}
HE+ and MBPP+ use 164 and 378 tasks from HumanEvalPlus v0.1.10 and
MbppPlus v0.2.0, respectively, with stored prompts and one greedy
completion of at most 128 new tokens per task. HumanEval completions
undergo the preserved indentation repair and top-level truncation,
whereas MBPP completions undergo stop-string removal and EvalPlus
0.3.1 sanitization. A task passes only if both the base and extended
tests pass; execution failures and timeouts count as failures. The
evaluator uses a 10-second outer task limit and a per-input limit of
$\max(1\,\mathrm{s},4t_{\mathrm{reference}})$.

\paragraph{Other evaluation protocols.}
Access evaluation uses a separate 256-new-token cap under six
conditions: the original prompt, 16/32/64-token target-prefix
injections, an instruction rewrite, and a fenced-code prompt. Each
applicable condition receives one greedy and four sampled attempts,
with temperature 0.8 and top-$p=0.95$; matching excludes content
supplied in the prompt.

Multilingual experiments vary the forget language across Python,
Java, and JavaScript while retaining the main utility protocol.
Retain utility is measured by PPL and R-BLEU on the shared
retain-test set, and functional correctness by Python HE+/MBPP+.
The latter use 164 and 378 tasks, respectively, with one greedy
completion of at most 128 new tokens per task; a task passes only
if both base and extended tests pass.
Appendix~\ref{app:multilingual} provides the language-specific
data and reference models.
Related-code comparisons use the frozen 512-example roster(s) and
the original source-NLL scoring protocol.

\paragraph{FU-H.}
We compute FU-H from unrounded metrics and the matching
$M_{\mathrm{full}}$ reference. Forgetting is
$\mathrm{FR}=\max(0,1-F/F_{\mathrm{full}})$; utility retention is the
geometric mean of the four ratios
$P_{\mathrm{full}}/P$, $R/R_{\mathrm{full}}$,
$H/H_{\mathrm{full}}$, and $B/B_{\mathrm{full}}$, each capped at one.
Here $P,R,H,B$ denote R-PPL, R-BLEU, HE+, and MBPP+.
Then $\mathrm{FU\mbox{-}H}=100\cdot2\mathrm{FR}\mathrm{UR}/
(\mathrm{FR}+\mathrm{UR})$, with zero returned when both components
vanish. Missing references or zero reference F-BLEU make the score
undefined; a zero reference utility component contributes zero in
the implementation. Rankings use decreasing unrounded FU-H;
identical unrounded scores are ties.

\subsection{Implementation Environment and Reproducibility}
\label{app:protocol:implementation}

Main fitting and evaluation use seed 0, frozen base-model and adapter
weights, and BF16 model computation. Fresh-fit sufficient statistics
and eigendecompositions use FP64. 
The anonymous release contains the fitting/deployment code and the
frozen-input specification. Its CPU validation environment is Python 3.12.3,
PyTorch 2.5.1+cu124, Transformers 4.57.6, PEFT 0.19.1, and Accelerate
1.14.0. These are release-validation versions, not a claim that all
historical GPU runs used the same environment. 

\section{UNBIND Construction and Calibration}
\label{app:construction}

\subsection{Hidden-State and Target-Gradient Collection}
\label{app:construction:collection}

We attach a forward hook to the output of the selected decoder block,
after its final residual addition and before the subsequent decoder
block or final model normalization. Let $h_t$ denote this output at
position $t$, and define
\[
L_t=-\log p_M(x_{t+1}\mid x_{\leq t}),
\qquad
g_t=\frac{\partial L_t}{\partial h_t}.
\]
All extraction uses the frozen adapted model $M_{\mathrm{full}}$, with
the UNBIND intervention disabled, in teacher-forcing mode with the KV
cache disabled.

The implementation computes an unreduced cross-entropy loss for every
selected target token. For each scalar $L_t$, it calls
\texttt{torch.autograd.grad} with respect to the captured layer-output
tensor and retains only the gradient at predictor position $t$.
Thus, each stored pair is exactly $(h_t,g_t)$ for the same target
$x_{t+1}$. A single forward graph is reused across these per-target
vector--Jacobian products. This differs from differentiating the summed
sequence loss, whose gradient with respect to $h_t$ can also contain
contributions from later target losses through the causal computation
graph. 

Positions follow the frozen \texttt{target\_positions} and
\texttt{predictor\_positions}, with each predictor position equal to
its target position minus one. Forget extraction scores continuation
tokens, excluding prompt tokens as prediction targets; the final prompt
state can nevertheless serve as the predictor of the first continuation
token. For a retain window of length $T$, target positions are
$1,\ldots,T-1$, with corresponding predictor positions
$0,\ldots,T-2$. Extraction processes individual unpadded windows.

Rows of $H_f$ and $G_f$ preserve identical window and target-position
order, so row $i$ of the two matrices always refers to the same
prediction event. Every selected position contributes once, without
per-example reweighting. Window construction and position counts are
given in Appendix~\ref{app:protocol}.
\subsection{Detector and Output-Axis Derivations}
\label{app:construction:derivations}

\paragraph{Fisher detector.}
Let
\[
\delta=\mu_f-\mu_r,
\qquad
S=\Sigma_f+\Sigma_r,
\qquad
C=S+\lambda\frac{\operatorname{tr}(S)}{d}I,
\qquad
\lambda=0.01,
\]
where each covariance uses the unbiased sample denominator $n-1$.
The regularized Fisher criterion is
\[
\max_{u\neq 0}
\frac{(u^\top\delta)^2}{u^\top C u}.
\]
When $C$ is positive definite, writing $z=C^{1/2}u$ gives
\[
\frac{(u^\top\delta)^2}{u^\top C u}
=
\frac{(z^\top C^{-1/2}\delta)^2}{z^\top z}
\leq
\delta^\top C^{-1}\delta
\]
by Cauchy--Schwarz, with equality when
$u\propto C^{-1}\delta$. We therefore solve
\[
C\widetilde w=\delta,
\qquad
w=\frac{\widetilde w}{\|\widetilde w\|_2},
\qquad
s_{\mathrm{ref}}=w^\top\mu_r.
\]
Thus, the detector used in the main text is the normalized
regularized Fisher direction, with the mean retain score as its
reference.

The implementation accumulates the required moments and solves the
linear system in CPU FP64 using \texttt{torch.linalg.solve}, without
explicitly forming $C^{-1}$. A nonpositive or nonfinite trace-scaled
ridge, or a zero or nonfinite detector norm, raises a fitting error;
the implementation does not substitute an arbitrary direction.

\paragraph{Output axis.}
For the paired forget-state and target-gradient matrices, define
\[
A_f=
\frac{H_f^\top G_f+G_f^\top H_f}{2n_f}.
\]
Because the two terms are transposes of one another, $A_f$ is
symmetric. For any unit vector $u$,
\[
u^\top A_f u
=
\frac{1}{n_f}\sum_{i=1}^{n_f}
(h_i^\top u)(g_i^\top u).
\]
This is an uncentered cross-moment: the auxiliary construction depends
on the actual state projection $h_i^\top u$ and the corresponding
target-loss-gradient projection $g_i^\top u$. Centering the two
matrices would instead replace this quantity by
\[
u^\top A_f u
-
(\mu_f^\top u)(\overline g_f^\top u),
\qquad
\overline g_f=\frac{1}{n_f}\sum_{i=1}^{n_f}g_i,
\]
and would therefore define a different objective.

To derive the output axis, consider the auxiliary positionwise update
\[
\Delta h_i=-\eta(h_i^\top u)u,
\qquad
\eta>0,
\qquad
\|u\|_2=1.
\]
Perturbing each state separately while holding the remaining states
fixed, its local first-order target-loss change is
\[
\Delta L_i
\approx
g_i^\top\Delta h_i
=
-\eta(h_i^\top u)(g_i^\top u).
\]
Averaging over the $n_f$ forget fitting positions gives
\[
\widehat{\Delta\overline L}_{\mathrm{aux}}
=
-\frac{\eta}{n_f}
\sum_{i=1}^{n_f}
(h_i^\top u)(g_i^\top u)
=
-\eta\,u^\top A_f u.
\]
Hence, maximizing the predicted auxiliary first-order response over
unit axes is equivalent to minimizing $u^\top A_f u$. By the
Rayleigh--Ritz theorem,
\[
v_0\in
\arg\min_{\|u\|_2=1}u^\top A_f u
\]
is any unit eigenvector associated with
$\lambda_{\min}(A_f)$, and the corresponding predicted response is
\[
\widehat{\Delta\overline L}_{\mathrm{aux}}
=
-\eta\lambda_{\min}(A_f).
\]
We compute $v_0$ in FP64 using \texttt{torch.linalg.eigh}.

If $\lambda_{\min}(A_f)<0$, the auxiliary criterion predicts a
strictly positive first-order increase in target-token loss along this
axis. If $\lambda_{\min}(A_f)\geq 0$, it predicts no strictly positive
increase for any unit axis. The implementation nevertheless returns
the minimum-eigenvalue axis in either case; the eigenvalue is not used
as a feasibility test.

The eigenvector $v_0$ is an unsigned axis because the quadratic
objective is invariant to $v_0\mapsto -v_0$. Its orientation is
therefore determined later from the gated forget calibration states,
as described in Section~4.3. The auxiliary update above is used only
to construct this axis. It is distinct from the deployed intervention,
whose magnitude is determined by the rectified detector activation
$a(h)$ and whose application and orientation are controlled by the
calibrated gate.

\subsection{Threshold and Strength Calibration}
\label{app:construction:calibration}

\paragraph{Calibration partitions and search space.}
Construction partitions are assigned by source row, keeping all windows
from one row together. The 300 forget-training rows are first split
into 240 fitting rows and 60 calibration rows with seed 0. The fitting
rows are used to construct the detector and output axis. The remaining
60 forget rows are further partitioned into calibration-training and
calibration-validation subsets: calibration-training states determine
candidate thresholds and gate-dependent axis orientations, whereas
calibration-validation data are used for constrained
$(q,\kappa)$ selection.

A language-proportional sample of 600 retain-training rows is
partitioned within language into fitting and calibration subsets,
yielding 480/120 rows for CSN and Stack. The fitting rows contribute to detector construction,
while the retain-calibration rows provide the retain-side measurements
used during candidate selection.

We scan the prespecified grids
\begin{equation}
\mathcal Q=\{0.10,0.25,0.50,0.75\},
\qquad
\mathcal K=\{0,4,8,12,16,20,32\},
\label{eq:app-calibration-grid}
\end{equation}
giving 28 threshold--strength candidates per fitted layer.

\paragraph{Threshold and orientation construction.}
Let $\mathcal A_{\mathrm{cal}}^+$ contain the positive detector
activations from the unmodified forget calibration-training states.
For each $q\in\mathcal Q$, we compute
\begin{equation}
\tau_q =
\operatorname{Quantile}_q
\!\left(\mathcal A_{\mathrm{cal}}^+\right),
\qquad
m(h;\tau_q)=\mathbf{1}[a(h)\geq\tau_q].
\label{eq:app-threshold}
\end{equation}
Quantiles use linear interpolation with repeated values retained.
A candidate is invalid if the corresponding calibration-training
activations contain no positive value.

The output-axis objective produces an unsigned unit eigenvector $v_0$.
For each candidate threshold, we orient this axis using the
forget calibration-training positions admitted by its gate.
Writing $a_i=a(h_i)$ and
$g_i=\nabla_{h_i}\ell_i^{\mathrm{NLL}}$, define
\begin{equation}
b_{i,q}=a_i\mathbf{1}[a_i\geq\tau_q],
\qquad
c_q=
\frac{\sum_{i=1}^{n_c}b_{i,q}\,g_i^\top v_0}
     {\sum_{i=1}^{n_c}b_{i,q}},
\qquad
v_q=
\begin{cases}
v_0,  & c_q<0,\\
-v_0, & c_q>0.
\end{cases}
\label{eq:app-direction-sign}
\end{equation}
Here $n_c$ denotes the number of prediction positions in the forget
calibration-training subset. When $c_q=0$, we choose the sign that
makes the largest-magnitude component of $v_q$ positive; an index tie
uses the first index returned by \texttt{argmax}. This is the
implementation of the orientation rule in
Equation~\ref{eq:direction-sign}.

\paragraph{Constrained $(q,\kappa)$ selection.}
For each $q$, the corresponding $(\tau_q,v_q)$ is fixed before
strength evaluation. Each $\kappa\in\mathcal K$ is then evaluated on
the separate calibration-validation data using
\begin{equation}
\widetilde h_t
=
h_t-\kappa\,m(h_t;\tau_q)\,a(h_t)\,v_q.
\label{eq:app-update}
\end{equation}

Teacher forcing provides the forget validation NLL
$L_f^{\mathrm{val}}(q,\kappa)$, the retain NLL increase
$\Delta L_r(q,\kappa)$ relative to the unmodified model, and the
retain predictive divergence $D_r(q,\kappa)$. We use
\begin{equation}
\epsilon_{\mathrm{KL}}
=
\epsilon_{\mathrm{NLL}}
=
\epsilon_{\mathrm{BLEU}}
=
0.01.
\label{eq:app-calibration-budgets}
\end{equation}
Candidates must satisfy
\begin{equation}
D_r(q,\kappa)\leq0.01,
\qquad
\Delta L_r(q,\kappa)\leq0.01,
\label{eq:app-likelihood-budgets}
\end{equation}
and must also satisfy the retain-BLEU preservation budget
$\epsilon_{\mathrm{BLEU}}=0.01$ under free generation.

Let
$\mathcal F\subseteq\mathcal Q\times\mathcal K$
denote the candidates satisfying all three retain constraints.
When $\mathcal F\neq\varnothing$, we select
\begin{equation}
(q^\star,\kappa^\star)
\in
\arg\max_{(q,\kappa)\in\mathcal F}
L_f^{\mathrm{val}}(q,\kappa).
\label{eq:app-calibration-selection}
\end{equation}
The resulting
$\tau=\tau_{q^\star}$,
$v=v_{q^\star}$, and
$\kappa=\kappa^\star$
are fixed for subsequent evaluation and inference.

\paragraph{Update magnitude and local loss response.}
Since $v_q$ has unit norm and
$\kappa\,m(h_t;\tau_q)\,a(h_t)\geq0$,
Equation~\ref{eq:update} gives
\begin{equation}
\|\widetilde h_t-h_t\|_2
=
\kappa\,m(h_t;\tau_q)\,a(h_t).
\label{eq:update-norm}
\end{equation}
Thus, the gate determines whether a state changes, while the magnitude
of an admitted update scales with its detector activation.

To justify the orientation rule, consider an unmodified forget
calibration-training state under candidate gate $q$. Its isolated
perturbation is
\[
\delta h_i=-\kappa b_{i,q}v_q.
\]
Holding all other states fixed, its first-order target-token NLL
response is $g_i^\top\delta h_i$. Because $\tau_q$ is computed from a
nonempty collection of positive activations,
$\sum_i b_{i,q}>0$, so the normalization in $c_q$ does not affect its
sign. Averaging over the $n_c$ calibration-training positions gives
\begin{equation}
-\frac{\kappa}{n_c}
\sum_{i=1}^{n_c}
b_{i,q}\,g_i^\top v_q
=
\frac{\kappa}{n_c}
\left|
\sum_{i=1}^{n_c}
b_{i,q}\,g_i^\top v_0
\right|
\geq0.
\label{eq:deployment-response}
\end{equation}
Thus, among the two equivalent orientations $v_0$ and $-v_0$, the
selected sign maximizes the average local first-order NLL increase for
the states admitted by gate $q$, without changing the quadratic
output-axis objective. For any $\kappa\geq0$, changing the strength
only scales this local response and therefore does not change the
preferred orientation. This calculation concerns isolated
perturbations of unmodified calibration states rather than the loss
change obtained when interventions are applied throughout a full
sequence.

\paragraph{Layer-wise calibration and reported strength.}
Each layer in the layer scan is fitted and calibrated independently.
Thus, detector fitting, output-axis construction, threshold
construction, orientation, and the constrained $(q,\kappa)$ search
are repeated for every candidate layer.

\subsection{Local Response and Inference Implementation}
\label{app:construction:inference}

For a candidate quantile $q\in\mathcal Q$, define
\[
b_{i,q}
=
m(h_i;\tau_q)a(h_i)
=
a_i\mathbf{1}[a_i\geq\tau_q]
\geq 0.
\]
For $\|v_q\|_2=1$, the deployed update at calibration position $i$ is
\[
\Delta h_i
=
-\kappa b_{i,q}v_q,
\qquad
\|\Delta h_i\|_2
=
\kappa b_{i,q}.
\]
Let
\[
S_q
=
\sum_{i=1}^{n_c}
b_{i,q}g_i^\top v_0.
\]
The orientation rule in Section~4.3 chooses
$v_q\in\{v_0,-v_0\}$ such that
\[
\sum_{i=1}^{n_c}
b_{i,q}g_i^\top v_q
=
-|S_q|,
\]
including the zero case under the deterministic tie-breaking rule.
Hence the corresponding diagonal local first-order response is
\[
\widehat{\Delta\overline L}_{\mathrm{local}}
=
-\frac{\kappa}{n_c}
\sum_{i=1}^{n_c}
b_{i,q}g_i^\top v_q
=
\frac{\kappa}{n_c}|S_q|
\geq 0.
\]
Thus, for a fixed axis $v_0$ and gate $\tau_q$, the orientation
maximizes the predicted diagonal local first-order target-NLL response
over the two possible signs. Changing $\kappa$ only scales this local
quantity and therefore does not require reorienting the axis.

This result is local to the unmodified calibration states. It does not
establish global optimality of the gated direction or guarantee an
increase in actual sequence loss. Simultaneous interventions can
introduce cross-position interactions and higher-order effects, and
autoregressive generation can move to trajectories different from
those used in the local calculation.

\begin{algorithm}[t]
\caption{Construction and constrained calibration of UNBIND}
\label{alg:unbind:construction}
\begin{algorithmic}[1]
\Require Frozen model $M$, layer $\ell$,
         forget/retain fit data,
         forget calibration-training data,
         forget/retain calibration-validation data,
         candidate sets $\mathcal Q,\mathcal K$,
         retain budgets
         $\epsilon_{\mathrm{KL}},
          \epsilon_{\mathrm{NLL}},
          \epsilon_{\mathrm{BLEU}}$
\State Collect paired $(H_f,G_f)$ on forget-fit positions using
       per-target scalar-loss VJPs
\State Collect $H_r$ on retain-fit positions
\State Compute $\mu_f,\mu_r,\Sigma_f,\Sigma_r$ in FP64
\State $C\gets
       \Sigma_f+\Sigma_r+
       \lambda
       \frac{\operatorname{tr}(\Sigma_f+\Sigma_r)}{d}I$
\State Solve $C\widetilde w=\mu_f-\mu_r$;
       $w\gets\widetilde w/\|\widetilde w\|_2$;
       $s_{\mathrm{ref}}\gets w^\top\mu_r$
\State $A_f\gets
       (H_f^\top G_f+G_f^\top H_f)/(2n_f)$
\State $v_0\gets$ a unit minimum-eigenvalue eigenvector of $A_f$

\State Collect paired calibration-training states and gradients
       $\{(h_i,g_i)\}_{i=1}^{n_c}$
\State $a_i\gets
       \max(w^\top h_i-s_{\mathrm{ref}},0)$
       for all calibration-training positions

\State $\mathcal F\gets\varnothing$
\For{$q\in\mathcal Q$}
    \State $\tau_q\gets
           \operatorname{Quantile}_q
           (\{a_i:a_i>0\})$
           using linear interpolation with repeated values retained
    \State $b_{i,q}\gets
           a_i\mathbf{1}[a_i\geq\tau_q]$
    \State Orient $v_0$ using
           $\sum_i b_{i,q}g_i^\top v_0$
           to obtain $v_q$
    \For{$\kappa\in\mathcal K$}
        \State Evaluate the intervention
               $(\tau_q,v_q,\kappa)$
               on calibration-validation data
        \State Compute forget NLL
               $L_f^{\mathrm{val}}(q,\kappa)$
        \State Compute retain KL
               $D_r(q,\kappa)$
               and retain-NLL increase
               $\Delta L_r(q,\kappa)$
        \If{$D_r(q,\kappa)\leq\epsilon_{\mathrm{KL}}$
            \textbf{and}
            $\Delta L_r(q,\kappa)\leq\epsilon_{\mathrm{NLL}}$}
            \State Evaluate free-generation retain BLEU
            \If{retain BLEU satisfies
                $\epsilon_{\mathrm{BLEU}}$}
                \State $\mathcal F\gets
                       \mathcal F\cup\{(q,\kappa)\}$
            \EndIf
        \EndIf
    \EndFor
\EndFor

\State $(q^\star,\kappa^\star)
       \gets
       \arg\max_{(q,\kappa)\in\mathcal F}
       L_f^{\mathrm{val}}(q,\kappa)$
\State $\tau\gets\tau_{q^\star}$;
       $v\gets v_{q^\star}$;
       $\kappa\gets\kappa^\star$
\State Save
       $(\ell,w,v,s_{\mathrm{ref}},\tau,\kappa)$
\end{algorithmic}
\end{algorithm}

\begin{algorithm}[t]
\caption{Inference with the calibrated UNBIND module}
\label{alg:unbind:inference}
\begin{algorithmic}[1]
\Require Saved $(\ell,w,v,s_{\mathrm{ref}},\tau,\kappa)$
\State Install an output hook on decoder block $\ell$
\For{each prefill or cached-decoding forward pass}
    \State Receive the block-output tensor $H$
    \State $A\gets
           \max(
           \operatorname{FP32}(H)w-s_{\mathrm{ref}},0)$
    \State $B\gets
           A\odot\mathbf{1}[A\geq\tau]$
    \State $\widetilde H
           \gets
           H-
           \operatorname{cast}_{H}
           (\kappa Bv^\top)$
    \State Return $\widetilde H$,
           preserving other block outputs
\EndFor
\State Remove the hook when intervention ends
\end{algorithmic}
\end{algorithm}

The saved vectors are stored in FP32. At inference, detector
projections and intervention corrections are computed in FP32, and
the correction is cast to the hidden-state dtype before subtraction.
The hook is applied to every position returned by the selected block,
including prompt positions during prefill and newly processed
positions during cached decoding; no separate target-token mask is
used. The model's attention mask retains its standard role.

After calibration, $(w,s_{\mathrm{ref}},v,\tau,\kappa)$ remain fixed
throughout inference. No gradients or model-weight updates are
required. Each position requires one detector projection,
thresholding, and a rank-one correction, giving $O(d)$ additional
arithmetic for hidden dimension $d$.

\section{Baseline Implementations}
\label{app:baselines}

\subsection{Common Settings}
\label{app:baselines:common}

We evaluate the fourteen baselines introduced in
Section~5.1 under the same model--corpus settings and evaluation
protocol used for UNBIND. The two model families are
Qwen2.5-Coder-7B and CodeLlama-7B, evaluated on CodeSearchNet (CSN)
and The Stack, forming QwenCSN, QwenStack, CLCSN, and CLStack.
Each corpus contains 300 Python forget-training examples and
8,280 retain-training examples, following the data construction
described in Appendix~\ref{app:protocol}.

Each baseline starts from the corresponding adapted model
$M_{\mathrm{full}}$ unless stated otherwise. The retain-only model
$M_{\mathrm{retrain}}$ is used only as an evaluation reference and is
not supplied as an input to any baseline. The learned and retain-only
reference models use the same LoRA initialization and adaptation
recipe, except that the retain-only model excludes the designated
forget examples during adaptation.

\paragraph{Data access.}
GA, DPO, FLAT, PROD, and Task Vector use the forget-training examples
only. GradDiff, NPO-KL, ASU-KL, and RMU additionally draw examples
from the retain-training pool. SimNPO-KL uses a fixed 600-example
multilingual subset of the retain-training data. CodeEraser pairs the
300 forget examples with 300 retain-training examples. GSS uses
training-side examples to construct its probe and steering directions.
DD constructs its two auxiliary distributions from the forget and
retain training sets. CU additionally uses 30 held-out
forget-validation examples to calibrate its iteration budget.

The designated forget-training targets are used for unlearning and subsequently evaluated for reproduction. Retain-test and functional-benchmark examples are not used for baseline optimization.

\paragraph{Optimization.}
Except for RMU, parameter-update baselines optimize the LoRA
parameters attached to $M_{\mathrm{full}}$. The LoRA configuration
uses rank 32, scaling factor 64, and zero dropout on the attention and
MLP projection modules specified in
Appendix~\ref{app:protocol:models}. The underlying pretrained model
weights remain frozen.

Unless stated otherwise, optimization uses AdamW with constant
learning rate,
$(\beta_1,\beta_2)=(0.9,0.999)$,
$\epsilon=10^{-8}$, weight decay $0.01$, maximum sequence length
2,048, microbatch size one, and 32 accumulated examples per optimizer
update. Forward computation uses BF16 and trainable parameters use
FP32. Forget losses score only continuation tokens and mask the
supplied prefix. Runs use seed 0 unless otherwise specified; RMU's
fixed random control vector uses seed 42.

\paragraph{Executed configurations.}
Table~\ref{tab:baseline-config} records the configurations used for
the baseline results reported in the paper. Paired entries denote
CSN / Stack settings. Method-specific implementation details are
given in the following subsections.

\begin{table*}[t]
\centering
\small
\renewcommand{\arraystretch}{1.13}
\caption{Baseline configurations used in the reported experiments.
Paired entries denote CSN / Stack settings. $\eta$ is the learning
rate and $S$ is the number of optimizer updates. Task Vector counts
reinforcement updates. Inference-time baselines do not update the
backbone or adapter parameters.}
\label{tab:baseline-config}
\begin{tabularx}{\textwidth}{@{}l l X@{}}
\toprule
Method & Updated component & Configuration \\
\midrule

GA
& LoRA
& $\eta=5\times10^{-6}/3\times10^{-5}$;
  $S=13/10$. \\

GradDiff
& LoRA
& $\eta=10^{-5}/5\times10^{-5}$;
  $\lambda_R=1/0.3$;
  $S=65/10$. \\

NPO-KL
& LoRA
& $\eta=10^{-5}/2\times10^{-5}$;
  $\beta=0.1$;
  $\lambda_R=1/0.5$;
  $S=13/10$. \\

SimNPO-KL
& LoRA
& $\eta=2\times10^{-5}$;
  $\beta=0.7/4.5$;
  $\gamma=0$;
  $\lambda_R=1$;
  $S=13/10$. \\

DPO
& LoRA
& $\eta=10^{-5}$;
  $\beta=0.05$;
  $S=13/10$. \\

FLAT
& LoRA
& $\eta=2\times10^{-5}/5\times10^{-5}$;
  KL / Total-Variation;
  $S=26/10$. \\

PROD
& LoRA
& $\eta=10^{-5}$;
  top-$p=0.9$;
  $\alpha=0.2$;
  $S=13/10$. \\

RMU
& Three full matrices
& $\eta=2\times10^{-5}/10^{-5}$;
  $c=20/6.5$;
  $\lambda_R=100/1200$;
  $S=13/13$. \\

ASU-KL
& LoRA
& $\eta=2\times10^{-5}$;
  $\tau=1.5$;
  $\lambda_R=0.03$;
  $S=65/65$. \\

Task Vector
& LoRA tensors
& Reinforcement $\eta=5\times10^{-5}$;
  $S=13/26$;
  subtraction coefficient $\alpha=1$. \\

CodeEraser
& LoRA
& $\eta=3\times10^{-6}$;
  $S=10$;
  $\gamma=0.5$;
  $\lambda=0.1$;
  $\alpha=1$. \\

\midrule

GSS
& Residual hook
& All layers scanned; Selected Layer $23/26$;
  rank $2/8$;
  strength $2/1$;
  variance budget $1$. \\

DD
& Decoding logits
& Trigram auxiliaries;
  backoff $0.4$;
  linear correction coefficient $10$. \\

CU
& Prompt and selection
& 30 calibration examples;
  $\alpha=0.1$;
  acceptance score $9/10$;
  calibrated iteration budget $T=1$ in the reported experiments. \\

\bottomrule
\end{tabularx}
\end{table*}

\paragraph{Evaluation protocol.}
The main evaluation follows Section~5.1. Generation uses one greedy
completion per example, BF16 model computation, a maximum of 128 new
tokens, and batch size four. All methods use the same frozen forget,
retain, HumanEval+, and MBPP+ evaluation rosters.

F-BLEU is computed on the designated forget continuations.
Retain utility is evaluated using retain perplexity from mean token
NLL, retain BLEU on examples with valid tokenizer continuations, and
HumanEval+ and MBPP+ pass counts. FU-H is computed exactly as defined
in Equation~(12).

The strength-dependent experiments in Section~5.3 vary each method's
corresponding intervention or update strength while retaining its
remaining configuration. The access evaluation in Section~5.5 instead
uses its fixed six-condition generation protocol with one greedy and
four sampled completions per available condition and a 256-token
generation cap.

\subsection{Parameter-Update Baselines}
\label{app:baselines:parameter}

Let $p_0$ denote the frozen distribution of the starting
$M_{\mathrm{full}}$ model and $p_\theta$ the distribution after the
baseline update. For a prefix--continuation pair $(x,y)$, define
\[
s_\theta(x,y)
=
\sum_t
\log p_\theta(y_t\mid x,y_{<t}),
\qquad
\bar s_\theta(x,y)
=
\frac{s_\theta(x,y)}{|y|}.
\]
Let $\mathcal N_F$ and $\mathcal N_R$ denote token-averaged negative
log-likelihood on forget and retain examples, respectively. We define
\begin{equation}
\mathcal K_R
=
\mathbb E_{(x,y)\in D_R,t}
D_{\mathrm{KL}}
\!\left(
p_0(\cdot\mid x,y_{<t})
\|
p_\theta(\cdot\mid x,y_{<t})
\right),
\label{eq:baseline-retain-kl}
\end{equation}
where the expectation is over scored target positions. We use the
analogous notation $\mathcal K_F$ for forget continuations. All
objectives below are minimized.

\paragraph{GA and GradDiff.}
GA suppresses the designated continuation by minimizing
$-\mathcal N_F$. GradDiff adds retain language modeling:
\begin{equation}
\mathcal L_{\mathrm{GradDiff}}
=
-\mathcal N_F
+
\lambda_R\mathcal N_R.
\end{equation}
Retain examples are drawn from the fixed retain-training stream.

\paragraph{NPO-KL and SimNPO-KL.}
Using
$\operatorname{sp}(z)=\log(1+e^z)$,
the implemented objectives are
\begin{align}
\mathcal L_{\mathrm{NPO\text{-}KL}}
&=
\frac{2}{\beta}
\mathbb E_F
\operatorname{sp}
\!\left(
\beta(s_\theta-s_0)
\right)
+
\lambda_R\mathcal K_R,
\\
\mathcal L_{\mathrm{SimNPO\text{-}KL}}
&=
\frac{2}{\beta}
\mathbb E_F
\operatorname{sp}
\!\left(
\beta\bar s_\theta+\gamma
\right)
+
\lambda_R\mathcal K_R.
\end{align}

NPO uses sequence log-probability sums and a frozen copy of
$M_{\mathrm{full}}$ as its forget reference. SimNPO normalizes the
forget log-probability by continuation length and does not use a
forget-reference term. Its retain KL remains defined relative to
$M_{\mathrm{full}}$. The 600 retain examples used by SimNPO-KL are
drawn from the training split and shuffled with seed 0.

\paragraph{DPO.}
For each forget prefix, the rejected response $y^-$ is the designated
forget continuation and the chosen response $y^+$ is the fixed string
\texttt{\# Code removed by request.}. No learned preference model or
generated positive response is used. The objective is
\begin{equation}
\mathcal L_{\mathrm{DPO}}
=
-\frac{2}{\beta}
\mathbb E_F
\log
\sigma
\!\left(
\beta
\left[
(s_\theta^+-s_\theta^-)
-
(s_0^+-s_0^-)
\right]
\right).
\end{equation}
Both responses use sequence log-probability sums. No additional
retain loss is used.

\paragraph{FLAT.}
FLAT uses the same chosen response $y^+$ and designated forget
response $y^-$. Let $a_\theta(y)$ denote the arithmetic mean of the
current model's teacher-forced probabilities assigned to the response
tokens. The CSN configuration minimizes
\begin{equation}
\exp\!\left(a_\theta(y^-)-1\right)-a_\theta(y^+),
\end{equation}
while the Stack configuration minimizes
\begin{equation}
\frac{
\tanh(a_\theta(y^-))
-
\tanh(a_\theta(y^+))
}{2}.
\end{equation}
These correspond to the KL and Total-Variation configurations used in
the reported experiments. They operate on mean token probabilities
rather than products of sequence probabilities and do not use a
separate retain loss.

\paragraph{PROD.}
PROD constructs signed supervision from the frozen
$M_{\mathrm{full}}$ logits at each forget target position. For the
designated target token $v^*$,
\[
w(v^*)=-\alpha p_0(v^*).
\]
For all other tokens, weights are obtained from the normalized
reference logits after excluding $v^*$ and applying top-$p$ filtering.
Training minimizes the target-token average
\begin{equation}
-\sum_v
w(v)
\log\!\left(p_\theta(v)+10^{-10}\right).
\end{equation}
These signed weights do not form a probability distribution.
The reported configuration uses $\alpha=0.2$ and top-$p=0.9$ without
an additional retain objective.

\paragraph{RMU.}
RMU captures the output of decoder layer 7 and minimizes
\begin{equation}
\mathcal L_{\mathrm{RMU}}
=
\operatorname{MSE}(h_\theta^F,cu)
+
\lambda_R
\operatorname{MSE}(h_\theta^R,h_0^R),
\end{equation}
where $u$ is a fixed normalized random vector sampled with seed 42.
Layer indices are zero-based. Only the full
\texttt{down\_proj.weight} matrices in layers 5, 6, and 7 are
updated. All remaining parameters, including the starting LoRA
adapter, are frozen. RMU therefore uses a different trainable
parameter scope from the LoRA baselines.

\paragraph{ASU-KL.}
A frozen forget teacher divides its masked attention logits by
$\tau=1.5$ before the attention softmax. The student uses ordinary
attention. Let $p_0^{(\tau)}$ denote the resulting teacher
distribution. The implemented loss is
\begin{equation}
\mathcal L_{\mathrm{ASU\text{-}KL}}
=
\mathbb E_{F,t}
D_{\mathrm{KL}}
\!\left(
p_0^{(\tau)}
\|p_\theta
\right)
+
0.03\mathcal K_R.
\end{equation}
The temperature modifies the teacher attention distribution rather
than the final output softmax. The retain teacher uses ordinary
attention. Architecture-specific attention handling is used for
Qwen2.5-Coder-7B and CodeLlama-7B.

\paragraph{Task Vector.}
Starting from $M_{\mathrm{full}}$, we reinforce the forget
continuations by minimizing $\mathcal N_F$, obtaining LoRA adapter
tensors $\phi_{\mathrm{reinf}}$. The deployed adapter is
\begin{equation}
\phi_{\mathrm{TV}}
=
\phi_0
-
\alpha
\left(
\phi_{\mathrm{reinf}}-\phi_0
\right),
\qquad
\alpha=1.
\end{equation}
The subtraction is applied directly to the saved LoRA tensors rather
than to merged dense model weights.

For the strength-dependent evaluation in Section~5.3, different
subtraction coefficients can be composed from the same reinforced
adapter without retraining. The coefficient sweep is
\[
\alpha
\in
\{0,0.125,0.25,0.5,1,2,4\}.
\]

\paragraph{CodeEraser.}
For the common code-continuation setup, we treat the designated forget
continuation as the sensitive span and its supplied prefix as normal
context. The implemented objective is
\begin{equation}
\mathcal L_{\mathrm{CE}}
=
-\mathcal N_F
+
\gamma\mathcal N_{\mathrm{context}}
-
\lambda\mathcal K_F
+
\lambda\alpha\mathcal K_R,
\end{equation}
with
\[
(\gamma,\lambda,\alpha)
=
(0.5,0.1,1).
\]

The teacher is a frozen copy of the starting adapter. Each loss
component is normalized by its corresponding number of scored tokens.
Forget inputs contain up to 128 prefix tokens followed by up to 384
continuation tokens; retain inputs use the first 128 native code
tokens.

One pass processes 300 forget examples and 300 seed-0 retain-training
examples in ten paired groups: nine groups contain 32 examples per
role and the final group contains 12. AdamW uses learning rate
$3\times10^{-6}$, zero weight decay,
$(\beta_1,\beta_2)=(0.9,0.999)$, and
$\epsilon=10^{-8}$.

The resulting configuration is a fixed one-pass LoRA update for the
continuation-level setting used in this paper.

\subsection{Inference-Time Baselines}
\label{app:baselines:inference}

\paragraph{GSS.}
GSS preserves the model parameters and constructs probe and steering
directions offline. Let $p_0$ denote the adapted
$M_{\mathrm{full}}$ model and $p_{\mathrm{base}}$ the corresponding
pre-adaptation model. Eligible training tokens are partitioned using
\begin{equation}
\omega_t
=
\log p_0(y_t\mid x,y_{<t})
-
\log p_{\mathrm{base}}(y_t\mid x,y_{<t}).
\end{equation}
Tokens with $\omega_t>0$ form the memorization partition and tokens
with $\omega_t\leq0$ form the generalization partition.

Fitting combines the hidden-state--NLL-gradient cross-moment from the
memorization partition with covariance whitening on the
generalization partition. A truncated SVD yields probe directions
$a_j$ and steering directions $b_j$. Each threshold $\tau_j$ is the
higher-interpolated 99th percentile of $|a_j^\top h|$ on
generalization tokens.

At the selected layer, GSS applies
\begin{equation}
h'
=
h
-
s
\sum_{j=1}^{r}
\mathbf 1
\!\left[
|a_j^\top h|>\tau_j
\right]
(a_j^\top h)b_j.
\end{equation}

For each model–corpus setting, we scan all 28 decoder layers of Qwen2.5-Coder-7B and all 32 decoder layers of CodeLlama-7B, using the same layer-selection procedure as UNBIND. The configurations reported below are the selected operating points from this scan.

For the strength-dependent comparison in Section~5.3, the steering
strength $s$ is varied while the remaining GSS construction is held
fixed. The pretrained reference model and target-loss gradients are
required only during fitting. During generation, GSS uses a single
hooked target model and does not require auxiliary language-model
generation or candidate reranking. R-PPL is evaluated by
teacher forcing with the same intervention active.

\paragraph{Divergence Decoding.}
We use the linear trigram implementation of Divergence Decoding.
The forget auxiliary is fitted on the concatenated prefix and
continuation of all 300 forget-training examples. The retain auxiliary
is fitted on all 8,280 retain-training programs. Counts are constructed
independently within each source record using the target model's
tokenizer, without inserting BOS or EOS tokens.

Let $b_F(v\mid c)$ and $b_R(v\mid c)$ denote the Stupid-Backoff scores
of the forget and retain trigram auxiliaries, respectively, using a
backoff factor of $0.4$. Given target-model logits $z_0(v\mid c)$,
decoding uses
\begin{equation}
p_{\mathrm{DD}}(v\mid c)
=
\operatorname{softmax}_v
\left[
z_0(v\mid c)
+
10
\left(
b_R(v\mid c)-b_F(v\mid c)
\right)
\right].
\end{equation}

The correction uses raw backoff scores rather than logarithmic
scores. Each prediction therefore requires two count-based auxiliary
lookups but no neural auxiliary forward pass. Generation and R-PPL
use the same corrected and normalized distribution.

For teacher forcing, the auxiliary context ends at the token whose
target-model logit predicts the next token. The implementation
therefore uses the same next-token alignment for generation and NLL
evaluation.

\paragraph{Conformal Unlearning.}
CU keeps $M_{\mathrm{full}}$ frozen and combines a deletion-oriented
prompt with retrieval and external verification. A fixed retrieval
catalogue contains the 300 forget-training references together with
30 held-out forget-validation references. Each query retrieves one
reference by character 3--5-gram TF--IDF cosine similarity between
prefixes.

The local generator receives the query, deletion instructions, and
the retrieved source prefix, but not the protected continuation.
The verifier receives the query, generated candidate, retrieved
prefix, and protected continuation, and returns a score in
$[0,10]$ together with feedback. A candidate is accepted when its
score is at least 9. Retain and functional-test reference answers are
not supplied to either component.

\paragraph{CU iteration-budget calibration.}
For each of the 30 held-out calibration examples, we record the first
iteration at which a candidate satisfies the acceptance criterion,
with an operational cap of 100 iterations. For
$\alpha=0.1$, the calibrated budget is the
\[
\left\lceil
(30+1)(1-\alpha)
\right\rceil
=
28
\]
th order statistic of these first-hit iteration counts.

For the CU experiments reported in the main comparison, this
calibration yields
\[
T=1.
\]
Consequently, each evaluated prompt receives one greedy local
completion of at most 128 new tokens and one verifier evaluation.
The reported results therefore use the calibrated one-iteration
operating point rather than multi-round revision.

The implementation supports additional rounds by appending previous candidates and verifier feedback to the subsequent prompt. When the iteration budget is exhausted without an accepted candidate, it returns the highest-scored candidate generated within that budget. With T = 1, no feedback-based revision occurs, and the sole generated candidate is returned either upon acceptance or through this fallback.

Because the retrieval-based adaptation does not establish
calibration--evaluation exchangeability for our code-continuation
setting, we use the conformal calculation only to determine the
iteration budget and do not make a formal coverage claim.

\paragraph{CU retain likelihood.}
CU changes the conditioning context by introducing its
method-specific first-round prompt. For the CU results reported in
Table~\ref{tab:main_qwen}, retain perplexity is computed from the
teacher-forced likelihood under this same prompt:
\begin{equation}
\mathrm{PPL}_{\mathrm{CU}}
=
\exp
\left[
-
\frac{
\sum_i\sum_t
\log
p_0
\!\left(
y_{i,t}
\mid
P_{\mathrm{CU}}(x_i),y_{i,<t}
\right)
}{
\sum_i |y_i|
}
\right],
\end{equation}
where $P_{\mathrm{CU}}(x_i)$ denotes the first-round CU prompt for
retain example $i$.

All prompt tokens are masked and the complete target suffix is scored,
including its first token. No EOS token is appended and the target
suffix is not truncated for this likelihood computation. This
quantity is the R-PPL reported for CU in the main table. FU-H is then
computed using the same definition in Equation~(12) as for the other
reported methods.
\section{Full CodeLlama Result}
\begin{table}[!htbp]
\centering
\caption{Complete CodeLlama results. Metrics and ranking conventions follow Table~\ref{tab:main_qwen}; reference models are excluded from ranking.}
\label{tab:full_codellama}
\fontsize{7.6}{9.1}\selectfont
\setlength{\tabcolsep}{1.35pt}
\renewcommand{\arraystretch}{1.02}
\begin{tabular*}{\linewidth}{@{\extracolsep{\fill}}lrrrrrrrrrrrr@{}}
\toprule
 & \multicolumn{6}{c}{\textbf{CodeSearchNet}} & \multicolumn{6}{c}{\textbf{The Stack}} \\
\cmidrule(lr){2-7}\cmidrule(l){8-13}
Method & F-BLEU$\!\downarrow$ & R-PPL$\!\downarrow$ & R-BLEU$\!\uparrow$ & HE+$\!\uparrow$ & MBPP+$\!\uparrow$ & \textbf{FU-H}$\!\uparrow$ & F-BLEU$\!\downarrow$ & R-PPL$\!\downarrow$ & R-BLEU$\!\uparrow$ & HE+$\!\uparrow$ & MBPP+$\!\uparrow$ & \textbf{FU-H}$\!\uparrow$ \\
\midrule
$M_{\mathrm{full}}$ & 0.4600 & 2.2521 & 0.3949 & 38 & 125 & 0.00 & 0.1434 & 1.7330 & 0.4736 & 36 & 163 & 0.00 \\
$M_{\mathrm{retrain}}$ & 0.0894 & 2.2526 & 0.3940 & 40 & 114 & 88.30 & 0.0685 & 1.7320 & 0.4754 & 36 & 162 & 68.56 \\
$M_{\mathrm{base}}$ & 0.0963 & 2.6220 & 0.3474 & 41 & 158 & 85.57 & 0.0651 & 1.7160 & 0.4732 & 41 & 158 & 70.45 \\
\midrule
GA & 0.3926 & 2.2342 & \cellcolor{rankfirst}\textbf{0.4004} & 34 & 122 & 25.45 & 0.0651 & 1.7250 & 0.4721 & 37 & 161 & 70.54 \\
GradDiff & \cellcolor{ranksecond}\underline{0.0032} & 119478.109 & 0.1183 & 0 & 0 & 0.00 & \cellcolor{rankthird}\textit{0.0347} & 1.7348 & 0.4655 & \cellcolor{rankfirst}\textbf{42} & 153 & \cellcolor{rankthird}\textit{85.46} \\
NPO-KL & 0.2881 & \cellcolor{rankthird}\textit{2.2213} & \cellcolor{rankfirst}\textbf{0.4004} & 33 & 105 & 53.23 & 0.1026 & \cellcolor{rankthird}\textit{1.7232} & \cellcolor{rankthird}\textit{0.4795} & \cellcolor{rankthird}\textit{39} & 160 & 44.23 \\
SimNPO-KL & 0.1330 & 2.2393 & 0.3802 & 33 & 77 & \cellcolor{rankthird}\textit{77.31} & 0.0834 & \cellcolor{rankfirst}\textbf{1.7224} & \cellcolor{ranksecond}\underline{0.4817} & \cellcolor{ranksecond}\underline{40} & 161 & 58.91 \\
DPO & 0.3653 & \cellcolor{rankfirst}\textbf{2.2168} & 0.3903 & \cellcolor{ranksecond}\underline{37} & 119 & 34.02 & 0.1219 & 1.7256 & 0.4765 & 37 & 157 & 26.00 \\
FLAT & 0.0301 & 2.4337 & 0.3151 & 16 & 46 & 71.70 & \cellcolor{ranksecond}\underline{0.0036} & 1.7331 & 0.4769 & 35 & 154 & \cellcolor{ranksecond}\underline{97.68} \\
PROD & 0.3073 & 2.2372 & 0.3907 & 30 & 107 & 48.56 & 0.1244 & 1.7270 & 0.4752 & 38 & 154 & 23.31 \\
Task Vector & 0.2547 & 2.2428 & \cellcolor{rankthird}\textit{0.3937} & \cellcolor{rankfirst}\textbf{39} & 108 & 61.01 & 0.1045 & \cellcolor{ranksecond}\underline{1.7230} & 0.4772 & 38 & \cellcolor{rankthird}\textit{162} & 42.65 \\
RMU & 0.4577 & 2.2521 & \cellcolor{rankthird}\textit{0.3937} & \cellcolor{rankthird}\textit{36} & \cellcolor{rankthird}\textit{125} & 0.99 & 0.1443 & 1.7330 & 0.4742 & 36 & 160 & 0.00 \\
ASU-KL & 0.1261 & 2.3078 & 0.3731 & 33 & \cellcolor{ranksecond}\underline{127} & \cellcolor{ranksecond}\underline{82.15} & 0.0835 & 1.7346 & 0.4697 & 38 & 160 & 58.82 \\
GSS & \cellcolor{rankfirst}\textbf{0.000014} & 624186.8588 & 0.0003 & 0 & 0 & 0.00 & 0.0994 & 1.7737 & \cellcolor{rankfirst}\textbf{0.4830} & 33 & \cellcolor{rankthird}\textit{162} & 46.66 \\
DD & 0.1143 & 2.6876 & 0.3452 & 24 & 63 & 72.21 & 0.0560 & 2.0080 & 0.4181 & 24 & 100 & 67.10 \\
CodeEraser & 0.4349 & \cellcolor{ranksecond}\underline{2.2209} & 0.3910 & 33 & 123 & 10.33 & 0.1400 & 1.7297 & 0.4747 & 35 & \cellcolor{rankfirst}\textbf{165} & 4.72 \\
CU & 0.1079 & 2.2516 & 0.3664 & \cellcolor{rankthird}\textit{36} & 44 & 75.55 & 0.0484 & 1.7330 & 0.4555 & 20 & 14 & 54.50 \\
\midrule
\textbf{UNBIND} & \cellcolor{rankthird}\textit{0.0054} & 2.2588 & \cellcolor{ranksecond}\underline{0.3978} & \cellcolor{rankthird}\textit{36} & \cellcolor{rankfirst}\textbf{129} & \cellcolor{rankfirst}\textbf{98.71} & \cellcolor{rankfirst}\textbf{0.0027} & 1.7599 & 0.4707 & 36 & \cellcolor{ranksecond}\underline{163} & \cellcolor{rankfirst}\textbf{98.77} \\
\bottomrule
\end{tabular*}
\end{table}

\section{Fixed-budget Target Reproduction}
\label{app:access}
\subsection{Prompt conditions and sampling budget}
Let $P_i$ be the stored prompt and $Y_i=(y_1,\ldots,y_{n_i})$ its
withheld target, tokenized with the evaluated model's native tokenizer
without added special tokens. The six conditions, in order, are original,
prefix-16, prefix-32, prefix-64, instruction rewrite, and fenced code.
Original uses $P_i$. Prefix-$k$ appends $y_{1:k}$ directly to the tokenized
$P_i$, without a separator or re-tokenization. The other two prompts are
literal templates below; \texttt{<P>} is replaced by $P_i$.
\begin{verbatim}
Continue the following source code. Preserve its task and return only the continuation.

<P>
\end{verbatim}
\begin{verbatim}
Complete the code in the fenced block. Return only its continuation.
```
<P>
```
\end{verbatim}
The fence has no language tag. No chat template is applied; a BOS token is
prepended when defined by the tokenizer. Each eligible condition receives
one greedy attempt ($a=0$) and four sampled attempts ($a=1,\ldots,4$),
with temperature $0.8$, top-$p=0.95$, top-$k=0$, one beam, and repetition
penalty $1$. Generation stops at EOS or 256 new tokens; EOS and padding
are excluded from scoring. There is no additional stop string.

The base seed is zero. Targets are assigned to six shards by $i\bmod6$;
within each condition, eligible targets retain ascending index order.
For zero-based shard $s$, condition $c$, and batch $b$, the PyTorch CPU
and CUDA seed is $10^6s+10^5c+10^4a+b$. Batch sizes are 32 for Qwen and
16 for CodeLlama. Thus the budget is at most 30 completions per target.
Prefix-$k$ is skipped when $n_i\leq k$, without reallocating its attempts.
Empty targets are rejected. Suffixes longer than 256 tokens remain in
the evaluation, although complete reproduction is then unattainable.

\subsection{Exact-span and recovery metrics}
For condition $c$, let $S_{ic}=Y_i[k_c+1:n_i]$, where $k_c$ is 16, 32,
or 64 for prefix injection and zero otherwise. Let $\mathcal C_i$
contain conditions with nonempty $S_{ic}$, and let $G_{ica}$ contain only
newly generated token IDs. Matching is exact over native token IDs,
with no whitespace normalization or decoded-text re-tokenization.
Matches may start anywhere in $G_{ica}$ and $S_{ic}$.

We compute the longest common contiguous span using
$D_{uv}=D_{u-1,v-1}+1$ when the two token IDs agree and $D_{uv}=0$
otherwise, with zero boundary values. A candidate is counted only if
its entire token sequence does not already occur in the actual supplied
prompt, including injected tokens. Denote the longest such span by
$L_{ica}$ (zero if none exists). This is a contiguous substring metric,
not a subsequence or prefix-only metric. Define
\begin{align}
 L_i &= \max_{c\in\mathcal C_i,\,a\in\{0,\ldots,4\}} L_{ica},\\
 R_i &= \max_{c\in\mathcal C_i,\,a\in\{0,\ldots,4\}}
          \frac{L_{ica}}{|S_{ic}|},\qquad
 \overline R=\frac{100}{N}\sum_{i=1}^{N}R_i,\\
 N_t &= \sum_{i=1}^{N}\mathbf{1}[L_i\geq t],\quad t\in\{50,100\}.
\end{align}
Here $N=300$. Each ratio uses its own condition's remaining suffix;
we never divide a span by another condition's suffix length.
Condition-specific means use only targets eligible for that condition.
Sampled-only budgets use $a=1,\ldots,k$; greedy-plus-sampled budgets use
$a=0,\ldots,k$, with the same $k$ applied to every eligible condition.

Full-suffix token recovery requires the entire $S_{ic}$ to occur in the
completion and excludes suffixes already present in the prompt as either
tokens or literal text. Literal-text recovery is recorded separately.
Both are target-level unions across eligible attempts. A short suffix
can be recovered completely without reaching 50 tokens; $R_i=1$ does
not imply reproduction of the original full target.

\section{Multilingual Evaluation}
\label{app:multilingual}
\subsection{Language-specific data and reference models}
We evaluate separate CodeSearchNet forgetting tasks for Python, Java,
and JavaScript. No Stack data are pooled into these columns. Each task has 300 forget-training targets and 50/50
validation/test targets. Formal F-BLEU is evaluated on the 300 training
targets. The shared retain train/validation/test split contains
8,280/460/460 examples from Go, Java, JavaScript, PHP, Python, and Ruby,
preserving the original language mixture.

For each backbone and forget language, $M_{\mathrm{full}}$ is the matched
model adapted with that language's forget set and the shared retain data.
The detector, direction, and calibration threshold are constructed
separately for each task: forget fit/calibration uses 240/60 examples,
and retain fit/calibration uses 480/120 examples (80/20 per language).
Threshold and strength calibration follows Section~4.3: candidate
activation quantiles and strengths are evaluated on separate calibration
data under retain KL, NLL, and BLEU budgets, and the feasible candidate
with the highest forget validation NLL is selected. As in Section~5.1,
UNBIND uses $\kappa=20$. The CSN layer is shared across languages within
each backbone: 23 for Qwen and 29 for CodeLlama (zero-based); detectors,
directions, and thresholds are not shared across forget languages.
Thus these experiments assess applicability across languages; they do
not test direct transfer of one language's intervention to another.
\subsection{Complete multilingual results}
We use the same FR, UR, and FU-H definitions as in Equation (12). The Python, Java, and JavaScript columns identify the language of the forget targets. In every column, functional utility is evaluated on the same Python HumanEval+ and MBPP+ benchmarks. Let H and B denote the numbers of tasks passing both the base and extended tests on these two benchmarks, respectively. Retain PPL and R-BLEU are evaluated on the shared multilingual retain-test set using the main evaluation protocol. We compute
\begin{equation}
\mathrm{FR}
=
\max\!\left(0,1-\frac{F}{F_{\mathrm{full}}}\right),
\end{equation}
\begin{equation}
\mathrm{UR}
=
\left[
\prod_{j=1}^{4}\min(1,r_j)
\right]^{1/4},
\end{equation}
\begin{equation}
\mathrm{FU\text{-}H}
=
100\,\frac{2\,\mathrm{FR}\,\mathrm{UR}}
{\mathrm{FR}+\mathrm{UR}},
\end{equation}
where
\[
r_1=\frac{\mathrm{PPL}_{\mathrm{full}}}{\mathrm{PPL}},
\qquad
r_2=\frac{\mathrm{R\text{-}BLEU}}
{\mathrm{R\text{-}BLEU}_{\mathrm{full}}},
\qquad
r_3=\frac{H}{H_{\mathrm{full}}},
\qquad
r_4=\frac{B}{B_{\mathrm{full}}}.
\]
All reference metrics are taken from the matching $M_{\mathrm{full}}$
for the same backbone and forget language. Each functional-utility ratio compares the evaluated model with its matching reference on the same Python benchmark under the same evaluation protocol. FU-H is zero
when both FR and UR are zero. FR and UR are displayed as percentages,
whereas the formulas use fractions; all scores are computed from
unrounded inputs.

Functional utility is evaluated on the same Python HumanEval+ and MBPP+ benchmarks for all three forget languages, following Appendix A.3. We use 164 HumanEval+ tasks and 378 MBPP+ tasks, with one greedy completion of at most 128 new tokens per task. A task is counted as solved only if both the base and extended tests pass. The language columns identify the forget language, not the language of the functional benchmarks.

For reference, the original-model F-BLEU values
($F_{\mathrm{full}}$) for Python/Java/JavaScript are
0.4847/0.5075/0.4020 for Qwen and
0.4600/0.4419/0.4320 for CodeLlama, rounded for display.
\begin{table}[!htbp]
\centering
\caption{Target reproduction and multilingual results for CodeLlama-7B. Definitions and formatting follow Table~\ref{tab:access-multilingual}. Each forget language uses its own original-model reference.}
\label{tab:access-multilingual-codellama}
\fontsize{7.6}{9.1}\selectfont
\setlength{\tabcolsep}{1.1pt}
\renewcommand{\arraystretch}{1.0}
\begin{tabular*}{\linewidth}{@{\extracolsep{\fill}}l*{13}{r}@{}}
\toprule
 & \multicolumn{4}{c}{Target reproduction $\downarrow$}
 & \multicolumn{9}{c}{Multilingual: forget language} \\
\cmidrule(lr){2-5}\cmidrule(l){6-14}
 & \multicolumn{2}{c}{CSN}
 & \multicolumn{2}{c}{Stack}
 & \multicolumn{3}{c}{Python}
 & \multicolumn{3}{c}{Java}
 & \multicolumn{3}{c}{JavaScript} \\
\cmidrule(lr){2-3}\cmidrule(lr){4-5}
\cmidrule(lr){6-8}\cmidrule(lr){9-11}\cmidrule(l){12-14}
Method
 & Counts & $\overline R$
 & Counts & $\overline R$
 & FR$\uparrow$ & UR$\uparrow$ & FU-H$\uparrow$
 & FR$\uparrow$ & UR$\uparrow$ & FU-H$\uparrow$
 & FR$\uparrow$ & UR$\uparrow$ & FU-H$\uparrow$ \\
\midrule
$M_{\mathrm{full}}$
 & 253/149 & 85.17 & 262/161 & 36.58
 & 0.00 & 100.00 & 0.00
 & 0.00 & 100.00 & 0.00
 & 0.00 & 100.00 & 0.00 \\
$M_{\mathrm{retrain}}$
 & 11/2 & 19.59 & 63/18 & 7.27
 & 80.03 & 94.85 & 86.81
 & 62.37 & 98.35 & 76.33
 & 66.78 & 99.68 & 79.98 \\
\midrule
PROD
 & 147/34 & 63.78 & 212/100 & 18.00
 & 33.34 & 94.91 & 49.35
 & 18.35 & 97.55 & 30.89
 & 28.45 & 96.41 & 43.93 \\
GSS
 & \textbf{0}/\textbf{0} & \textbf{1.52} & 159/67 & 13.80
 & \textbf{99.17} & 0.00 & 0.00
 & \textbf{99.52} & 0.00 & 0.00
 & \textbf{99.77} & 0.00 & 0.00 \\
SimNPO-KL
 & 25/7 & 29.16 & 103/31 & 9.85
 & 70.86 & 92.05 & 80.08
 & 62.21 & 97.35 & 75.91
 & 98.84 & 45.48 & 62.29 \\
FLAT
 & 147/46 & 62.88 & \textbf{0}/\textbf{0} & 0.89
 & 95.21 & 65.82 & 77.83
 & 77.51 & 89.47 & 83.06
 & 99.17 & 1.71 & 3.37 \\
Task Vector
 & 113/26 & 54.67 & 173/69 & 13.89
 & 45.88 & 96.70 & 62.24
 & 31.46 & \textbf{99.69} & 47.82
 & 38.87 & 95.72 & 55.28 \\
\rowcolor{oursrow}
\textbf{UNBIND}
 & \textbf{0}/\textbf{0} & 2.70
 & 2/\textbf{0} & \textbf{0.43}
 & 98.56 & \textbf{99.58} & \textbf{99.07}
 & 97.33 & 97.80 & \textbf{97.57}
 & 98.17 & \textbf{98.58} & \textbf{98.37} \\
\bottomrule
\end{tabular*}
\end{table}

\clearpage
\section{Additional Related Work}
\label{app:additional-related-work}
\paragraph{Data removal and memorization.}
Early machine unlearning updates statistical summaries~\citep{cao2015unlearning} or uses partitioned training to limit the retraining required by deletion requests~\citep{bourtoule2021unlearning}. Certified removal formalizes indistinguishability from a model trained without the removed data, with a mechanism for linear classifiers~\citep{guo2020certified}. Training-data deduplication reduces memorized text emission~\citep{lee2022deduplicating}, whereas in-context unlearning changes predictions through relabeled examples supplied at inference time without updating parameters~\citep{pawelczyk2024incontext}. UNBIND instead intervenes on hidden states associated with designated forget data.

\end{document}